\documentclass[10pt, conference, letterpaper]{IEEEtran}
\usepackage{enumitem}
\usepackage{cite}
\usepackage{xspace}
\usepackage{amsmath,amssymb,amsfonts}
\usepackage{algorithmic}
\usepackage{algorithm}
\usepackage{array}
\usepackage[caption=false,font=footnotesize,labelfont=rm,textfont=rm]{subfig}
\usepackage{textcomp}
\usepackage{stfloats}
\usepackage{url}
\usepackage{verbatim}
\usepackage{listings}
\usepackage[dvipsnames]{xcolor}
\usepackage{float}
\usepackage{graphicx}
\usepackage{booktabs}
\usepackage{multirow}
\def \quark {\textit{Quark}\xspace}
\usepackage{amsthm}
\usepackage{hyperref}

\usepackage{tabularx}

\theoremstyle{plain}
\newtheorem{theorem}{Theorem}
\DeclareMathOperator*{\argmax}{argmax}

\def\BibTeX{{\rm B\kern-.05em{\sc i\kern-.025em b}\kern-.08em
    T\kern-.1667em\lower.7ex\hbox{E}\kern-.125emX}}

\begin{document}

\lstdefinestyle{python}{ 
    language=Python,
    basicstyle=\ttfamily\footnotesize,
    keywordstyle=\color{NavyBlue}\bfseries,
    commentstyle=\color{green!60!black},
    numbers=left,
    numberstyle=\tiny\color{gray},
    stepnumber=1,
    numbersep=5pt,
    backgroundcolor=\color{white},
    showspaces=false,
    showstringspaces=false,
    showtabs=false,
    frame=single,
    rulecolor=\color{black},
    tabsize=2,
    captionpos=b,
    breaklines=true,
    breakatwhitespace=false,
    title=\lstname,
    xleftmargin=10pt,
    framexleftmargin=10pt,
    emph={@staticmethod,self}, 
    emphstyle={\bfseries\color{Rhodamine}}, 
    escapeinside={\%*}{*)},
    morekeywords={action, int},
}

\lstdefinestyle{C}{ 
    language=C,
    basicstyle=\ttfamily\footnotesize,
    keywordstyle=\color{NavyBlue}\bfseries,
    commentstyle=\color{green!60!black},
    numbers=left,
    numberstyle=\tiny\color{gray},
    stepnumber=1,
    numbersep=5pt,
    backgroundcolor=\color{white},
    showspaces=false,
    showstringspaces=false,
    showtabs=false,
    frame=single,
    rulecolor=\color{black},
    tabsize=2,
    captionpos=b,
    breaklines=true,
    breakatwhitespace=false,
    title=\lstname,
    xleftmargin=10pt,
    framexleftmargin=10pt,
    escapeinside={\%*}{*)},
    morekeywords={action, int},
}

\title{\quark: Implementing Convolutional Neural Networks Entirely on Programmable Data Plane\thanks{Corresponding author: Lin Cui. Email: tcuilin@jnu.edu.cn}}  

\author{  
Mai Zhang\textsuperscript{1}, 
Lin Cui\textsuperscript{1, *}, 
Xiaoquan Zhang\textsuperscript{1}, 
Fung Po Tso\textsuperscript{2}, 
Zhen Zhang\textsuperscript{1}, 
Yuhui Deng\textsuperscript{1} 
and Zhetao Li\textsuperscript{1} \\
\textsuperscript{1}\textit{Department of Computer Science, Jinan University, Guangzhou, China.} \\
\textsuperscript{2}\textit{Department of Computer Science, Loughborough University, UK.} \\
}
\maketitle
\maketitle

\begin{abstract}
The rapid development of programmable network devices and the widespread use of machine learning (ML) in networking have facilitated efficient research into intelligent data plane (IDP). Offloading ML to programmable data plane (PDP) enables quick analysis and responses to network traffic dynamics, and efficient management of network links.
However, PDP hardware pipeline has significant resource limitations. For instance, Intel Tofino ASIC has only 10Mb SRAM in each stage, and lacks support for multiplication, division and floating-point operations. These constraints significantly hinder the development of IDP.
This paper presents \quark, a framework that fully offloads convolutional neural network (CNN) inference onto PDP. \quark employs model pruning to simplify the CNN model, and uses quantization to support floating-point operations. Additionally, \quark divides the CNN into smaller units to improve resource utilization on the PDP. 
We have implemented a testbed prototype of \quark on both P4 hardware switch (Intel Tofino ASIC) and software switch (i.e., BMv2). Extensive evaluation results demonstrate that \quark achieves 97.3\% accuracy in anomaly detection task while using only 22.7\% of the SRAM resources on the Intel Tofino ASIC switch, completing inference tasks at line rate with an average latency of 42.66$\mu s$.
\end{abstract}

\section{Introduction}
The development of programmable data plane (PDP) \cite{pdpsurvey} has enabled deep programmability for computer networks, allowing for flexible packet processing and forwarding at line rate. 
Concurrently, the rapid advancement of machine learning (ML) in networking and the gradual adoption of PDP devices have driven the evolution of the intelligent data plane (IDP) \cite{IDPs_survey}. By deploying ML models on PDP, IDP leverages both the line rate processing capabilities of PDP along with the intelligent analytical abilities of ML. 

Existing research has demonstrated the feasibility of IDP, primarily in areas like flow classification \cite{N3IC,swamy2022taurus,BNN2}, network defense \cite{FlowLens}, congestion control \cite{congestion}, and network telemetry \cite{MLpdpSurvey}. Some efforts have been made to offload various types of simple ML models onto PDP, such as decision trees\cite{DT1,DT2}, binary neural networks\cite{BNN1,N3IC,BNN2}.
Other efforts have focused on enhancing PDP with additional hardware (e.g., FPGA) in PDP's pipeline \cite{iisy2,zheng2022iisy,swamy2022taurus} to support ML functions. 
Convolutional neural networks (CNNs) \cite{cnnSurvey2}
, a cornerstone of deep learning, offer significant advantages in tasks like time series prediction and signal identification due to their ability to detect sequential correlations. However, offloading CNNs to PDP presents unique challenges.

Current attempts to implement CNNs on PDP have faced limitations. Some approaches only provide software switch implementations \cite{INQ-MLT,NetNN}, while others require additional devices such as CPUs \cite{banana}. These solutions fall short of fully utilizing PDP hardware, such as Tofino ASIC switches, thus introducing additional latency and compromising line-rate performance.

\textit{Fully implementing CNNs on PDP hardware switches is necessary but challenging}. For PDP hardware, such as switches with PISA (Protocol-independent switch architecture) \cite{Tofino_PISA}, which is a well-known hardware pipeline architecture in Intel Tofino ASIC, implementing CNNs on the hardware pipeline has the following challenges: 
\textit{(i)} The lack of support for floating-point operations in PISA, which are essential for CNNs, limits the effective offloading of these models. While high-precision floating-point computation has been achieved on PDP \cite{float_cal}, it exhausts PDP resources, making it unsuitable for CNN offloading. 
\textit{(ii)} The absence of multiplication and division support in the PISA architecture hinders the implementation of convolution and activation operations. Although multiplication can be implemented using bit shifts and addition, this approach consumes excessive stage resources (e.g., an 8-bit multiplication requires 4 stages)
\textit{(iii)} PISA’s pipeline-based operation can be decomposed into only a limited number of stages (e.g., Intel Tofino 1 offers 12 stages), each with restricted computational and storage capacities. Additionally, the pipeline does not support looping, further complicating CNNs deployment on PDP.

To address these challenges, we propose \quark (\underline{Q}uantized convol\underline{u}tional neur\underline{a}l netwo\underline{rk}), a framework for offloading CNN inference onto PDP while ensuring high inference accuracy, reducing computational demands, and minimizing forwarding latency.
\quark employs quantization techniques to convert network models into fixed-point numbers, overcoming the lack of floating-point support in PDP. \quark also uses model pruning techniques to eliminate unnecessary neurons or channels in the CNN, mitigating PDP resource constraints and reducing the number of recirculations, thereby reducing inference latency. Moreover, \quark achieves multiplication within one stage using match-actions in SRAM. Besides, \quark proposes a modular design by abstracting CNN model into multiple units of varying granularity, and maximizes the use of pipeline resources by combining different units to implement various CNN models. Specifically, our contributions are:

\begin{itemize}
    \item We propose the \quark framework, which completely offloads CNN inference onto PDP, maximizes resource utilization and achieves line-rate CNN inference on hardware switches.
    \item We utilize pruning and quantization on control plane to compress CNN models to overcome the limitations of hardware resources and floating-point computations. We also propose a modular design to split and reorganize CNN models to maximize the utilization of PDP resources.
    \item We have implemented a testbed prototype of \quark based on the Intel Tofino ASIC switches. Experimental results demonstrate that \quark effectively detects anomalies with an accuracy of 97.3\%, and performs inference tasks at line rate with 42.66$\mu s$ latency.
\end{itemize}


\section{Background and Related Works}

\subsection{Intelligent Data Plane}
With the increasing integration of ML in networking, IDP aims to offload various ML models onto PDP such as PISA switches. Currently, neural network-based IDP focuses primarily on Binary Neural Networks (BNNs) \cite{N3IC,BNN1,BNN2}, Recurrent Neural Networks (RNNs) \cite{RNN}, and Multi-Layer Perceptrons (MLPs) \cite{INQ-MLT,in3}, contributing to tasks such as traffic classification \cite{N3IC,swamy2022taurus,BNN2,iisy2}, network defense \cite{defense,FlowLens}, congestion control \cite{congestion,swamy2022taurus}, and load balancing \cite{congestion}. 

Beyond these models, CNNs excel at extracting time-series features from data flows, proving particularly effective for network functions, especially for network traffic intrusion detection\cite{intrusion_detection} and traffic classification \cite{traffic_classification}. However, the complex computations involved in CNN have limited their deployment on PISA hardware switches.

To maintain parallel processing capabilities, PISA's available operations and memory are highly restricted \cite{zhang2023dapper}. Various solutions have been employed to encode complex ML computations into basic PISA-supported operations. Typically, there are three main approaches:
\textit{(i)} Converting floating-point parameters to fixed-point numbers by using quantization techniques to transform operations into PISA-supported integer computations.
\textit{(ii)} Expanding the data plane by adding extra hardware modules to handle PISA's resource constraints.
\textit{(iii)} Implementing ML models on FPGA \cite{FPGA} or Network Processor \cite{NP}, as these platforms support operations unavailable in PISA, such as multiplication, division, and looping.

\subsection{Related Work}
Table \ref{tab:comparison} summarizes existing works on implementing neural networks on PDP. N2Net \cite{BNN1} and N3IC \cite{N3IC} have deployed BNNs on PDP to address the lack of floating-point computation support. BNNs convert the trained parameters of neural networks to 0 or 1, allowing logic operations to approximate multiplication and use integer-based approximations of floating-point \cite{BNNsurvey}. N3IC applies this approach to traffic analysis and deploys it on SmartNICs. 
However, the binarization of network parameters affects inference accuracy.

Besides binarization, other works have verified the feasibility of IDP on software switches (e.g., BMv2 \cite{bmv2}). NetNN \cite{NetNN} addresses the processing and memory constraints of the PISA architecture by partitioning deep neural networks (DNNs) across multiple switches on a layer-by-layer basis. 
INQ-MLT \cite{INQ-MLT} employs quantization techniques to convert floating-point operations into integer computations supported by PDP, addressing the lack of floating-point support in PISA. The system also uses Quantization Aware Training (QAT)\cite{QAT} to mitigate accuracy loss caused by quantization.

In addition, some works use additional hardware devices to assist PDP in achieving ML functionality. Taurus \cite{swamy2022taurus} extends PISA by incorporating a custom hardware MapReduce block between two stages in the pipeline. The blocks, emulated using FPGA, work alongside parsers and match-action tables (MATs) to handle packet forwarding. BaNaNa Split \cite{banana} also uses quantization, splitting the neural network model, with one part processed by the server CPU and the other part quantized and integrated into programmable network devices.




\begin{figure*}[]
    \setlength{\abovecaptionskip}{0pt}
    \setlength{\belowcaptionskip}{0pt}
    \centering
    \includegraphics[width=0.9\linewidth]{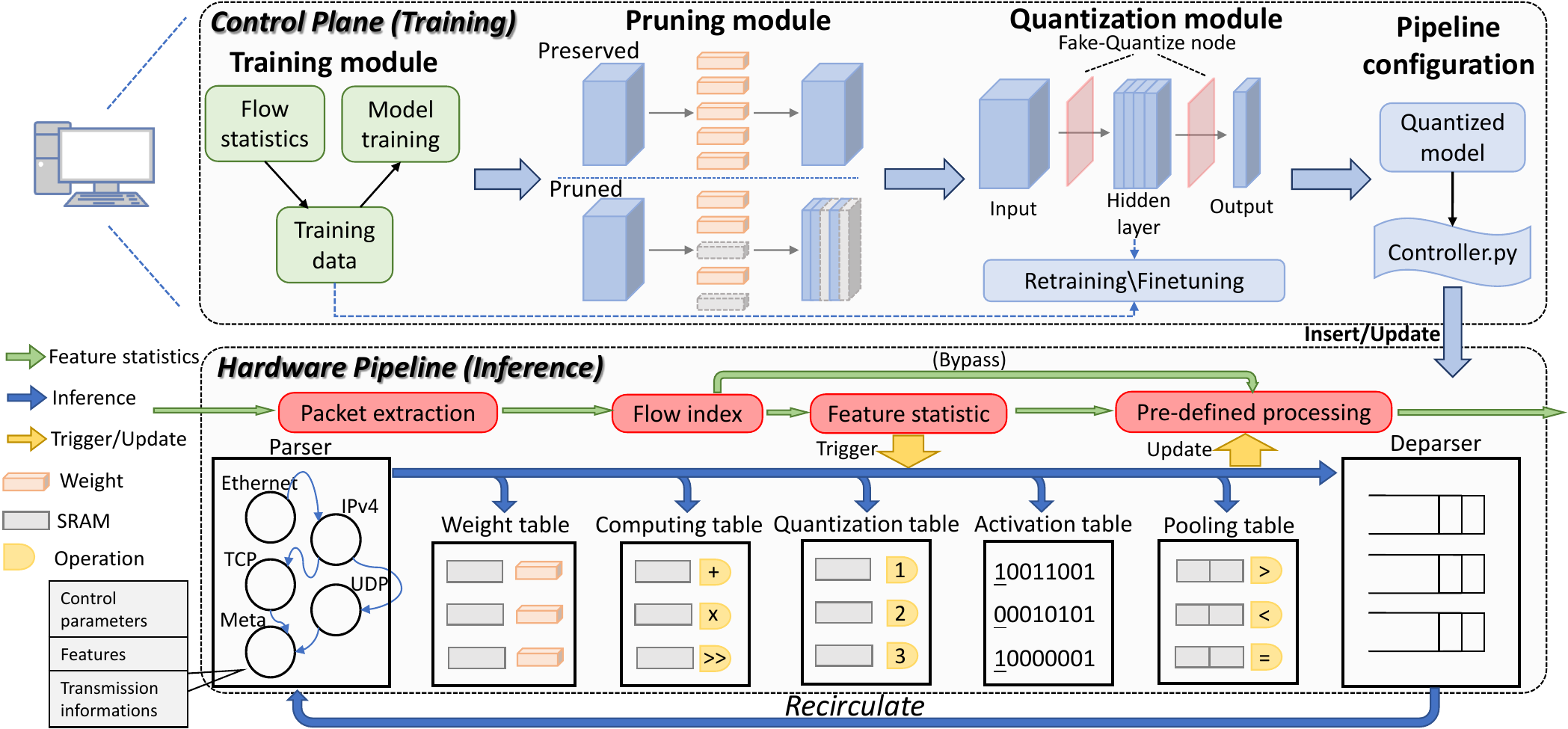}
    \caption{An overview of the \quark architecture.}
    \label{fig:overview}
    \vspace{-15pt}
\end{figure*}

\begin{table}[tb]
\caption{State of the art.}
\centering
\begin{tabular}{|>{\raggedright\arraybackslash}m{1.2cm}|>{\raggedright\arraybackslash}m{1.0cm}|>{\raggedright\arraybackslash}m{2.4cm}|>{\raggedright\arraybackslash}m{2.5cm}|}
\hline
\textbf{Work} & \textbf{Model} & \textbf{Target Platform} & \textbf{Notes} \\ \hline
N2Net\cite{BNN1} & BNN & Unknown Switches & Precision degradation (Binary) \\ \hline
N3IC\cite{N3IC} & BNN & NetFPGA & Precision degradation (Binary) \\ \hline
Taurus\cite{swamy2022taurus} & DNN & Intel Tofino ASIC with FPGA & Need additional Hardware Block \\ \hline
INQ-MLT\cite{INQ-MLT} & CNN \& MLP & BMv2 (software) & Excessive model complexity \\ \hline
BaNaNa Split\cite{banana} & NN & SmartNICs/Unknown Switches with CPU & High latency between PDP and CPU \\ \hline
NetNN\cite{NetNN} & DNN & BMv2 (software) & High cost and latency by multiple switches\\ \hline
\quark (ours) & CNN & Intel Tofino ASIC (hardware) \& BMv2 (software) & Entirely on hardware PDP, low latency and high accuracy \\ \hline
\end{tabular}
\label{tab:comparison}
\vspace{-5mm}
\end{table}

\subsection{Discussion}
Although binarization effectively addresses PDP floating-point constraints, it results in notable accuracy loss. Using quantization instead of binarization mitigates this drawback but introduces higher computational demands, requiring further optimization. Schemes deployed on software switches have also proposed methods to address PDP resource limitations, such as using TCAM to store dot product results. However, software switches have much lower processing efficiency compared to hardware switches, and the stricter resource limitations on hardware switches hinder the further development of these schemes. Using extra hardware (e.g., FPGA) to assist the hardware switch introduces inevitable delay between the hardware and PDP. This approach has lower scalability and flexibility and incurs additional costs. In this context, IN3\cite{in3} presents a novel architecture that integrates model compression with data plane pipelines for neural network inference.

Our objective is to design a solution that simplifies CNN models and fully offloads them onto the data planes of programmable hardware devices (e.g., Intel Tofino ASIC switch) without modifying existing switch hardware. This approach aims to address the resource and stage limitations of the PISA architecture, achieving efficient inference performance while maintaining high accuracy and low latency.

\section{Overview of \quark}
As shown in Fig. \ref{fig:overview}, the architecture of \quark contains two main components: the control plane and the hardware pipeline (i.e., data plane). 

\subsection{Control Plane}
The primary task of the control plane is to train a CNN model and insert or update its parameters into the hardware pipeline. The workflow is as follows:
\textit{(i)} Model Training: The model training module analyzes flow statistics to obtain training data and trains a floating-point model.
\textit{(ii)} Pruning: The pruning module performs channel pruning to remove less important parameter channels from the pre-pruned model. 
\textit{(iii)} Quantization: The quantization module employs QAT by adding fake-quantize nodes to the pruned model to simulate precision loss during quantized inference. The model is then retrained or fine-tuned for maintaining accuracy (see Section \ref{sec:compression}).
\textit{(iv)} Pipeline configuration: the pipeline configuration extracts the model parameters and uses the controller's API (e.g., P4Runtime) to insert or update them into the pipeline.

This architecture design enables efficient deployment of CNN models on PDP, ensuring high accuracy while reducing computational and storage resource requirements.

\subsection{Hardware Pipeline}
The hardware pipeline has two distinct workflows: network flow feature statistics (green arrow) and neural network inference (blue arrow), as shown in Fig. \ref{fig:overview}. 
\subsubsection{Feature Statistics} This workflow extracts network flow feature statistics (see Section \ref{sec:pipeline}).
\textit{(i)} When the switch receives a normal packet, it extracts the header information and retrieves the corresponding data via a flow index (e.g., whether the flow has already been predicted). 
\textit{(ii)} If prediction information is already indexed, feature statistics are skipped, and predefined processing is executed.
\textit{(iii)} If the prediction information is not indexed, feature statistics is performed, and the packet is forwarded.
\textit{(iv)} Once certain conditions are met during feature statistics (e.g., counting the first n packets), the neural network inference workflow is triggered.

\subsubsection{Neural Network Inference} In this workflow, we design a CNN inference model composed of multiple basic operations such as parser, deparser, MATs, and ALUs. 
\textit{(i)} The modular design (see Section \ref{sec:modular-CNN}) splits the CNN inference model into several basic units. 
\textit{(ii)} If the pipeline cannot fit the entire inference model, packets are cloned and recirculated within the pipeline to achieve complex calculations like convolution operations. This includes the weight table for retrieving weights, the computing table for handling the lack of multiplication operations in PDP, the quantization table for quantizing convolution results, the activation table for performing activation layer operations, and the pooling table for pooling operations. Parsers and deparsers handle packet header parsing and execute recirculate operations.
\textit{(iii)} Once inference is completed, the pipeline updates the prediction results in the flow index, which serves as the entry for subsequent packets.

\section{Model Compression for \quark}
\label{sec:compression}

\subsection{Pruning Module}
\quark employs model pruning techniques to simplify neural network structures. Given the high programmability and parallel computing capabilities of the PISA, efficient use of computational resources is crucial. Pruning drastically reduces the number of parameters while preserving accuracy \cite{pruning}. 
Specifically, \quark evaluates the importance of weights to identify and remove channels that minimally contribute to the model's prediction (see Fig. \ref{fig:overview}). This method achieves significant compression with minimal accuracy loss, prevents overfitting, and enhances the model's generalization ability and stability. Evaluation results demonstrate that the pruned CNN model of \quark exhibits higher computational efficiency and lower resource consumption on PDP while preserving high prediction accuracy.



\subsection{Quantizing CNN Parameters for \quark}
Due to the lack of support for floating-point operations within the pipeline, \quark employs a quantization module to convert the parameters of convolutional and fully connected layers in CNNs from floating-point precision (e.g., Float32) to lower-precision fixed-point integers (e.g., INT8). Fixed-point calculations can be performed within the PISA pipeline using basic operations, significantly reducing the complexity of neural network inference without substantial accuracy loss. 

We use \( r \) to denote the floating-point numbers and \( q \) to denote the quantized fixed-point integers.
The quantization process maps the parameter \( r \) from the continuous range \([r_{min}, r_{max}]\) to \( q \) within the range \([q_{min}, q_{max}]\). The range bounds $q_{min}$ and $q_{max}$ depend on the bit-width \( b \) of the integer \( q \), as well as whether the integer is signed or unsigned.
\begingroup
\setlength{\abovedisplayskip}{6pt}
\setlength{\belowdisplayskip}{6pt}
\begin{equation}
    \left ( {q}_{min},{q}_{max}\right )=\begin{cases}\left(-{2}^{b-1},{2}^{b-1}-1\right),&signed
\\ \left(0,{2}^{b}-1\right),&unsigned\end{cases} \,.
\end{equation}
\endgroup

Two key definitions for \quark are \(S\) (scale) and \(Z\) (zero-point). The scale \(S\) is the factor that converts floating-point numbers to fixed-point numbers while preserving their relative precision and proportion. The definition of \(S\) is:
\begin{equation}
\label{Scale}
S = \frac{r_{max} - r_{min}}{q_{max} - q_{min}} \,\,.
\end{equation}
The zero-point \(Z\) represents the integer value corresponding to zero in floating-point numbers after quantization, ensuring consistency of relative relationships between numbers before and after quantization. The definition of $Z$ is:

\begingroup
\setlength{\abovedisplayskip}{4pt}
\setlength{\belowdisplayskip}{4pt}
\begin{equation}
\label{Zeropoint}
Z = \text{Round}(q_{max} - \frac{R_{max}}{S})\,.
\end{equation}
\endgroup

These definitions allow \quark to convert from floating-point numbers to fixed-point integers:
\begingroup
\begin{equation}
\label{quanti_common}
    r=S(q-Z)\,,
\end{equation}
\endgroup
\begingroup
\begin{equation}
\label{quanti_rtoq}
q = \text{Clamp}(\text{Round}(\frac{r}{S}+Z),q_{min},q_{max})\,.
\end{equation}
\endgroup
The function of Clamp is to constrain the quantized fixed-point integer \( q \) within the fixed-point range \([q_{min}, q_{max}]\). Thus, \quark can use Equation \eqref{quanti_rtoq} to complete the quantization of parameters such as weights and biases.

\subsection{Quantizing Convolution for \quark}
In CNN, both convolutional layers and fully connected layers essentially perform matrix multiplications. Suppose \( r_1 \) and \( r_2 \) are two \( N \times N \) matrices in floating-point representation, and \( r_3 \) is the matrix resulting from their multiplication. \({r}_{3}^{i,j}\) is the element in the \(i\)-th row and \(j\)-th column of matrix \(r_3\).
\begingroup
\setlength{\abovedisplayskip}{5pt}
\setlength{\belowdisplayskip}{5pt}
\begin{equation}
\label{matrix_multiple}
{r}_{3}^{i,j}=\displaystyle\sum_{k=1}^{N}{r}_{1}^{i,k}\cdotp {r}_{2}^{k,j}\,.
\end{equation}
\endgroup


The convolution principle of CNN refers to performing a matrix multiplication between the weight matrix \(w\) and the input matrix \(x\), and adding the bias matrix \(b\) to obtain the output matrix \(a\). By substituting the above parameters into Equation \eqref{matrix_multiple} and simplifying the matrix notation, we obtain: 
\begingroup
\setlength{\abovedisplayskip}{5pt}
\setlength{\belowdisplayskip}{5pt}
\begin{equation}
    a=\displaystyle\sum_{i=1}^{N}{w}_{i}{x}_{i}+b \,.
\end{equation}
\endgroup
Assuming \( S_x, Z_x \) are the scale and zero-point for the input matrix \( x \), and similarly, \( S_w, Z_w \) for \( w \) and \( S_a, Z_a \) for \( a \). Additionally, \(q_x\), \(q_w\), and \(q_a\) represent the fixed-point matrices corresponding to matrices \(x\), \(w\), and \(a\), respectively. According to the Equation \eqref{quanti_common}, we can derive:
\begingroup
\setlength{\abovedisplayskip}{5pt}
\setlength{\belowdisplayskip}{5pt}
\begin{equation}
{S}_{a}({q}_{a}-{Z}_{a})\!=\! 
\displaystyle\sum_{i}^{N}{S}_{w}({q}_{w}\!-\!{Z}_{w}){S}_{x}({q}_{x}\!-\!{Z}_{x}) 
\!+\!{S}_{b}({q}_{b}\!-\!{Z}_{b})\,,
\end{equation}
\begin{equation}
\label{quanti_before}
{q}_{a}=\frac{{S}_{w}{S}_{x}}{{S}_{a}}\displaystyle\sum_{i}^{N}({q}_{w}-{Z}_{w})
({q}_{x}-{Z}_{x})+\frac{{S}_{b}}{{S}_{a}}({q}_{b}-{Z}_{b})+{Z}_{a} \,.
\end{equation}
\endgroup
Here, the non-integer parts are limited to \( {S}_{w}{S}_{x}\) and \({S}_{b}\). \( S \) and \( Z \) serve as intermediaries for converting between floating-point numbers and fixed-point integers, which only need to ensure that the conversion between $r$ and $q$ is reversible.
Therefore, we can use \( {S}_{w}{S}_{x}\) to replace \({S}_{b}\), and set \({Z}_{b}\) directly to zero. Thus, the Equation \eqref{quanti_before} can be adjusted to:
\begingroup
\setlength{\abovedisplayskip}{5pt}
\setlength{\belowdisplayskip}{5pt}
\begin{equation}
\label{quanti_final}
{q}_{a}=M(\displaystyle\sum_{i}^{N}({q}_{w}-{Z}_{w}) 
({q}_{x}-{Z}_{x})+{q}_{b})+{Z}_{a}\,,
\end{equation}
\endgroup
where 
\begingroup
\setlength{\abovedisplayskip}{5pt}
\setlength{\belowdisplayskip}{5pt}
\begin{equation}
\label{M_cal}
M = \frac{S_w S_x}{S_a} \,.
\end{equation}
\endgroup

Equation \eqref{M_cal} can be approximated as a fixed-point number with a bit shift on PISA hardware pipeline. These equations effectively implement quantized convolution operations within CNN on PDP which lack support for floating-point operations, ensuring both computational efficiency and accuracy.

\subsection{Quantization Aware Training}
The initial quantization parameters of the CNN was achieved using Equation \eqref{quanti_rtoq} to calculate directly from floating-point numbers. However, without retraining the model post-quantization, the model parameters' accuracy may be affected.

To address this issue, \quark utilizes Quantization Aware Training (QAT), an optimization technique that simulates the quantization process during training, thereby improving the model's performance in deployment \cite{QAT}.

In QAT, fake-quantize nodes are inserted into the network model (see Fig.\ref{fig:overview}). During forward propagation, QAT simulates the Clamp and Round operations of quantization, helping the model adapt to quantization effects and achieve higher accuracy. 
In backward propagation, the straight-through estimator (STE) \cite{ste} approximates the gradient of the quantization operation, as the Round function has zero gradient. This allows the gradient to propagate back to the weights before the fake-quantize nodes. As a result, the weights undergo pseudo-quantization, simulating quantization errors, and the gradients of these errors are propagated back to update the original weights, allowing the model to adapt to quantization.

\begin{figure}[t]
\centering
\includegraphics[width=3.5in]{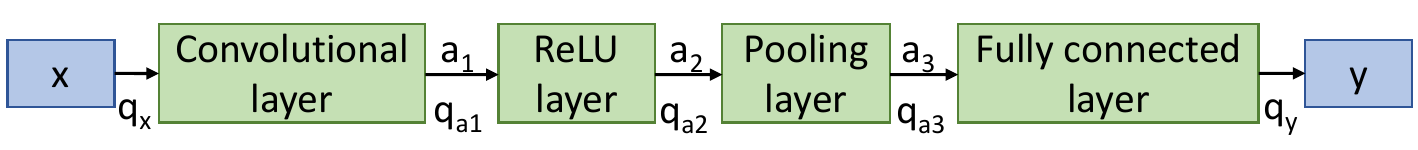}
\caption{A example of Quantized CNN model.}
\label{smallModel}
\vspace{-15pt}
\end{figure}
\subsection{Quantizing CNN Training and Inference}
We use an example CNN model in Fig. \ref{smallModel} to illustrate the CNN quantization process of \quark during forward propagation. Let the input and output be \( x \) and \( y \), with the outputs of the hidden layers denoted as \( a \). The quantized fixed-point numbers corresponding to these parameters are \( q_x \), \( q_y \), and \( q_a \).

During training, forward passes with sufficient data are performed to record the value ranges \([r_{min}, r_{max}]\) for \( x \), \( a_1 \), \( a_2 \), \( a_3 \), and \( y \). At the end of training, these recorded ranges are used to pre-calculate certain elements (e.g., $S$, \( M \), \( Z_x \), and \( Z_w \)) using Equations \eqref{Scale}, \eqref{Zeropoint} and \eqref{M_cal}.


During inference, \quark first quantize the input \( x \) to a fixed-point integer \( q_x \), and then use Equation \eqref{quanti_final} to compute the result of the first layer \( q_{a1} \). This process continues through the ReLU and Pooling layers. For fully connected layers, the quantization process is similar to that of convolutional layers, allowing us to compute \( q_y \) and obtain the final prediction.

This quantization method is scalable to larger networks due to its consistent underlying principles, and therefore establishing a robust quantization approach for CNN inference.

\section{\quark Design in Data Plane}
This section details the design and deployment of a quantized CNN on the data plane of a PISA hardware switch.

\subsection{Unit-based Modularization of CNN}
\label{sec:modular-CNN}

To fit a CNN inference model into a PDP hardware pipeline, we propose a modular design that abstracts and splits a CNN model into multiple units of varying granularity. 
For example, activation and pooling layers, with relatively low computational complexity, can be combined with convolutional layers into a single unit (see Section \ref{sec:CAP} for an example). Similarly, fully connected layers and activation layers can be merged into another unit. Different units can be defined by varying scales, allowing each of them to process one or multiple features. 
This flexibly allows \quark to tailor the CNN units to fit within the constraints of the available pipeline stages.

A CNN inference model can be achieved through the combination of one or multiple units. 
In PISA-based programmable devices, the data planes of different hardware devices offer varying amounts of resource space. We recommend deploying as many units as possible within a single pipeline. When necessary, the recirculation technique can be employed to force data packets to enter the pipeline multiple times for repeated unit processing. 
This modular design maximizes the utilization of the data plane’s resources, completes CNN inference with minimal latency, and provides a flexible and scalable solution for deploying CNNs on the PDP.

\begin{figure}[t]
\centering
\includegraphics[width=0.8\linewidth]{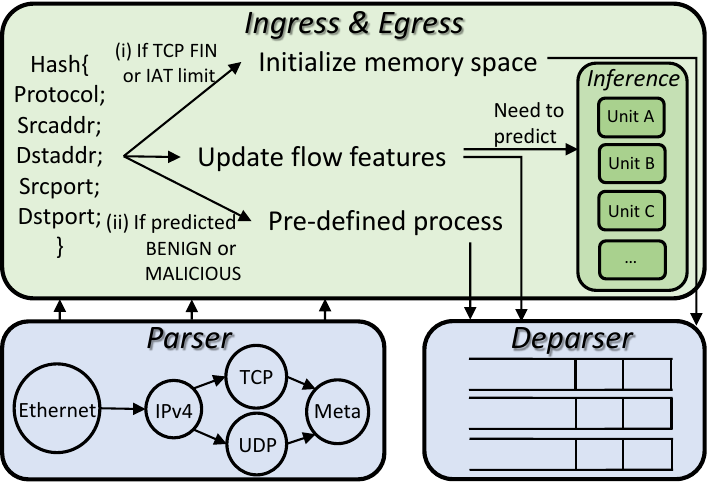}
\caption{Packet processing pipeline.}
\label{fig:pipeline}
\vspace{-12pt}
\end{figure}

\subsection{Quantized CNN Design in Pipeline}
\label{sec:pipeline}
We have implemented \quark on both P4 hardware (Intel Tofino ASIC) switches and BMv2. 
The processing pipeline of Quantized CNN in the PDP is illustrated in Fig. \ref{fig:pipeline}.

\textbf{Parser}: The parser extracts features from the packet header (e.g., packet length, TCP flags, IAT). These features play various roles in the ingress stage based on the flow's state.

\textbf{Ingress \& Egress}: Packets are hashed into the corresponding flow and processed as follows: \textit{(i)} If the packet header carries flags such as TCP FIN flag or exceeds the IAT limit, the corresponding flow's memory space is initialized; \textit{(ii)} If the flow matches pre-defined operations, it is classified according to the pre-defined results; \textit{(iii)} The flow's features are updated. If the packet triggers a specific condition (e.g., the arrival of the n-th packet), neural network inference is triggered, and the inference result is updated in the pre-defined results.

\textbf{Parameters}: Since biases and weights are fixed after training and quantization, storing them in match-action tables (MATs) is efficient. 
Due to the pipeline's characteristics, metadata is initialized to zero during recirculation, so \quark stores the layer outputs in the header. Additionally, parameters to control the inference process are required (see Table \ref{tab:variables}).

\begin{table}[b]
    \vspace{-15pt}
    \centering
    \caption{Parameter Description Table.}
    \begin{tabular}{|>{\raggedright\arraybackslash}m{1.5cm}|>{\raggedright\arraybackslash}m{6.5cm}|}
        \hline
        \textbf{Parameter} & \textbf{Description} \\
        \hline
        layer\_index & The current inferencing layer. All features reading and storage depend on this variable. \\
        \hline
        channel\_index & The channel currently being computed. It manages data storage and reading together with the \textit{layer\_index}. \\
        \hline
        conv\_flag & Whether accumulation is completed to determine if activation and pooling operations can be performed. \\
        \hline
        input\_index & Which two sets of features in the input matrix are currently being computed. \\
        \hline
    \end{tabular}
    \label{tab:variables}
\end{table}

\subsection{Implementation of CNN in the P4 Switch}
\label{sec:CAP}
\begin{figure}[t]
    \centering
    \includegraphics[width=0.7\linewidth]{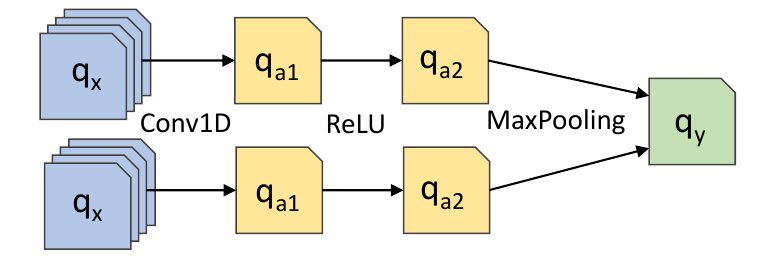}
    \vspace{-3pt}
    \caption{The structure of a CAP-Unit.}
    \label{fig:small_module}
    \vspace{-10pt}
\end{figure}

For simplicity, we use one-dimensional CNN (1D-CNN) for this introduction, with similar concepts extendable to multi-dimensional CNNs. A 1D-CNN unit, shown in Fig. \ref{fig:small_module}, is implemented on the Intel Tofino ASIC switch. Given the simplicity of activation and pooling layer computations, requiring only one stage each, we combined convolutional/fully connected, activation, and pooling layers into a single unit, termed as CAP-Unit (\underline{C}onvolutional, \underline{A}ctivation and \underline{P}ooling Layer \underline{Unit}). To maximize pipeline resource utilization, we designed each CAP-Unit to process two features simultaneously. 
Given Tofino's 12-stage limit, one CAP-Unit is deployed per pipeline, with the recirculate technique used to reprocess inference packets and complete the full CNN inference.

\begin{figure}[tbp]
    \centering
    \begin{lstlisting}[style=C, caption={P4$_{16}$ code of CAP-Unit.}, label={P4_code}]
apply {    
    /* (i) Retrieve the inputs from the header */
    get_input_tbl.apply();
    /* (ii) Retrieve the weights and biases */
    weight_tbl.apply(); bias_tbl.apply();    
    /* (iii) Computing two sets of features separately and storing the accumulated convolution results */
    multi_tbl_1.apply(); multi_tbl_2.apply();
    hdr.hdr_meta.acc_temp_1 += multi_result_1;
    hdr.hdr_meta.acc_temp_2 += multi_result_2;    
    /* When all channels have been accumulated, proceed with steps (iv)~(vii) */
    if (conv_flag == current_channel_num) {
       /* (iv) Quantize accumulation results */
       hdr.hdr_meta.acc_temp_1 += bias;
       hdr.hdr_meta.acc_temp_2 += bias; 
       quanti_tbl_1.apply(); quanti_tbl_2.apply(); 
       /* (v) ReLU module */
       if (result_1 < 0) { result_1 = 0; }
       if (result_2 < 0) { result_2 = 0; }        
       /* Skip (vi) when fully connected layer, do
          result_1 += result_2; */
       /* (vi) Maxpooling module */
       maxPooling_tbl.apply();        
       /* (vii) Storage */
       if (layer_index == 1) {
          storage_tbl_1.apply();
          /* Restart next Accumulation*/
          conv_flag=0; channel_index++;
          /* When finish current input features*/
          if(channel_index > current_channel_num){
             input_index++; channel_index = 0;}
          /* When finish current layer*/
          if (input_index > features_num) { 
             layer_index++; input_index = 0;}           
       }else if(layer_index == 2){
          //... }}
    /* recirculate */
    //...
}
    \end{lstlisting}
    \vspace{-25pt}
\end{figure}

\begin{figure}[t]
    \centering
    \includegraphics[width=1\linewidth]{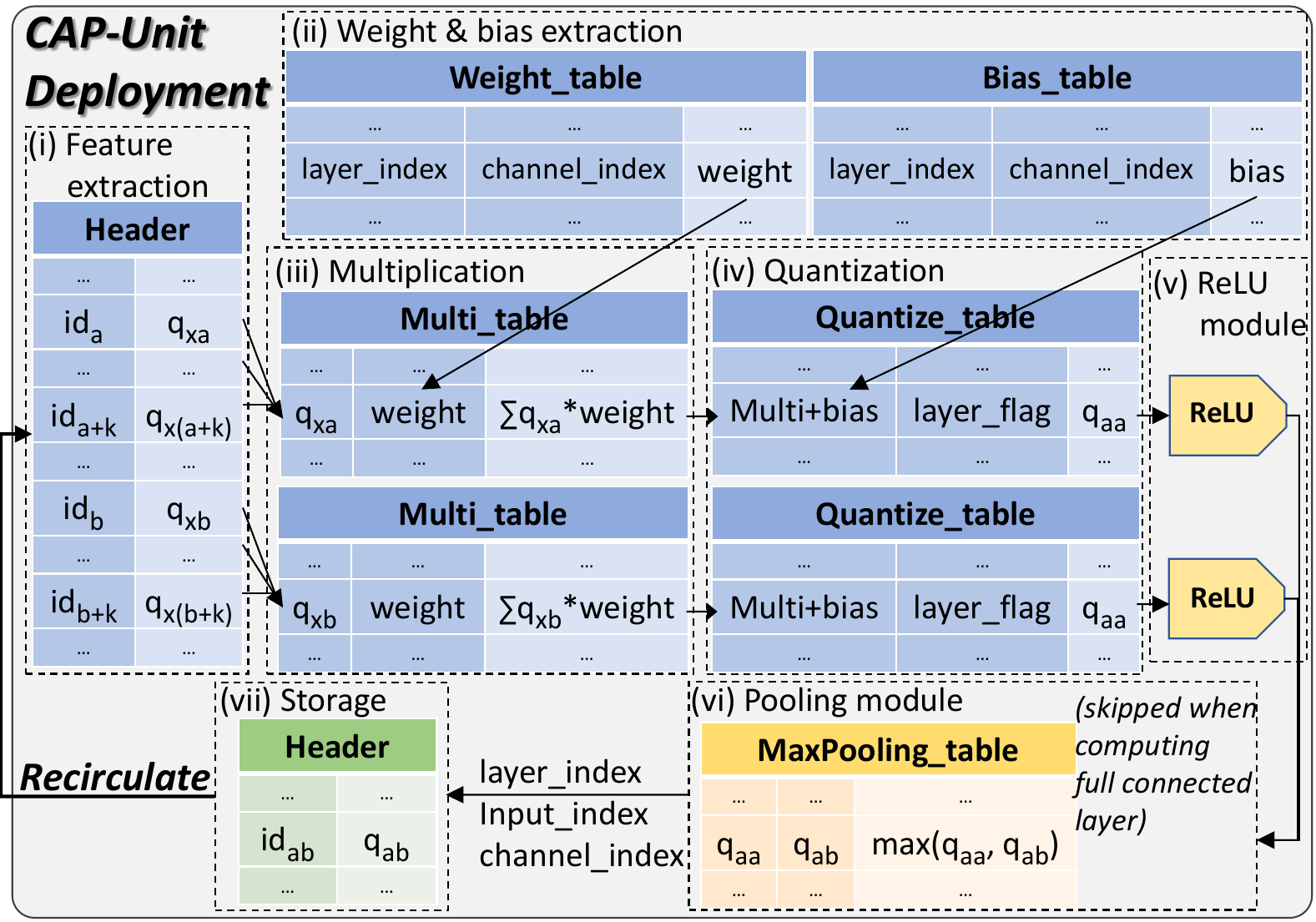}
    \caption{Deployment of the CAP-Unit in P4 pipeline.}
    \label{fig:deployment}
    \vspace{-15pt}
\end{figure}

Listing \ref{P4_code} presents the general logic of P4\(_{16}\) code of our design. 
We divide the computation process of the CAP-Unit into seven main steps (see Fig. \ref{fig:deployment}). Initially, \quark extracts the intermediate and control parameters in the header using the parser.
(\textit{i}) Using the \textit{layer\_index} and \textit{input\_index}, the input \({q}_{x}\) corresponding to the current layer 
are obtained and sent to the next stage.
(\textit{ii}) Based on the \textit{layer\_index}, the weights and biases of the current layer are extracted from the MATs.
(\textit{iii}) Since multiplication is not supported in the Tofino switch, a multiplication MAT is designed to perform the calculations. Although bit-shift multiplication is possible, it requires excessive pipeline resources (e.g., an 8-bit multiplication requires 4 stages). Instead, \quark stores all multiplication results in a MAT to reduce the number of required stages.
(\textit{iv}) According to Equation \eqref{quanti_final}, after completing the \(\displaystyle\sum\left({q}_{w}-{Z}_{w}\right) \left({q}_{x}-{Z}_{x}\right)\) operation, the accumulation result adds the bias, multiplies by \(M\), and adds the zero-point to complete the quantization.
Since floating-point operations are not supported in PISA and \(M\) remains constant during inference, \quark pre-compute the product of \(M\) and the previous results, storing them in a MAT to utilize relatively abundant SRAM resources and improve computation efficiency.
(\textit{v}) \quark uses ReLU for the activation layer, setting the convolution result to zero if it is less than zero, thus introducing non-linearity into our CNN. 
(\textit{vi}) For the pooling layer, \quark uses maxpooling to extract features and reduce dimensionality by selecting the maximum value in each region, retaining the most significant features while reducing the computational load. This step is skipped when computing the fully connected layers. 
(\textit{vii}) \quark stores the results into the header, controlled by \textit{layer\_index}, \textit{input\_index}, and \textit{channel\_index}. The storage method for \quark ensures that inputs are fully utilized before being overwritten, and avoids dirty data reads (see Section \ref{sec:header_bit} for more details). If all computations for the current layer are complete, \quark modify the \textit{layer\_index} to proceed to the next layer.

Additionally, the output of a single CAP-Unit corresponds to the output of a single channel in the CNN. Since hidden layers in CNN typically consist of multiple channels, the final output of a hidden layer is the sum of the outputs from each channel. To facilitate this, \quark utilize a temporary variable in the header to store the accumulated results from multiple CAP-Units during the recirculation operation (e.g., \textit{hdr\_meta.acc\_temp\_1} and \textit{hdr\_meta.acc\_temp\_2} in Listing \ref{P4_code}).

\begin{table}[b!]
\vspace{-15pt}
\caption{Symbols and descriptions.}
\centering
\begin{tabular}{|>{\raggedright\arraybackslash}p{0.8cm}|>{\raggedright\arraybackslash}p{7.2cm}|}
\hline
\textbf{Symbol} & \textbf{Description} \\ \hline
$T$ & Number of input features to the first layer \\ \hline
$L_{conv}$ & Number of convolutional layers \\ \hline
$L_{fc}$ & Number of fully connected layers \\ \hline
$C_{in}^{(n)}$ & Number of input channels for the $n$-th convolutional layer \\ \hline
$C_{out}^{(n)}$ & Number of output channels for the $n$-th convolutional layer \\ \hline
$T_{in}^{(m)}$ & Number of input features for the $m$-th fully connected layer \\ \hline
$T_{out}^{(m)}$ & Number of output features for the $m$-th fully connected layer \\ \hline
\end{tabular}
\label{table:definitions}
\end{table}

\subsection{Discussion and Analysis}
\subsubsection{Number of Recirculations}
With the modular design, if the PDP pipeline can not accommodate all units of a CNN model due to resource constraints, recirculations will be required. We analyze the upper bound of the number of recirculations when the CAP-Unit is used for the modularization. Table \ref{table:definitions} defines the symbols.



\begin{theorem}
When deploying a CNN model on PDP pipeline using the CAP-Unit, the maximum recirculations required is 
$\lceil(\!T \!+ \!L_{conv}\! +\! L_{fc}) \cdot C^2\rceil$, 
where $C\! =\! \max \limits_{1\leq n \leq L_{conv} , 1\leq m \leq L_{fc} }\{C_{in}^{(n)}, \!C_{out}^{(n)}, \!T_{in}^{(m)}, \!T_{out}^{(m)}\}$. 
\end{theorem}
\renewcommand\qedsymbol{$\blacksquare$}
\begin{proof}
Let $U$ denote the total number of CAP-Units of a CNN model:
\begingroup
\setlength{\abovedisplayskip}{4pt}
\setlength{\belowdisplayskip}{4pt}
\[ U\!=\!\sum_{n=1}^{L_{conv}} \!C_{in}^{(n)}\! \cdot\! C_{out}^{(n)}\! \cdot \lceil{T} /{2^{n}}\rceil\!+ \!\sum_{m=1}^{L_{fc}} T_{out}^{(m)} \!\cdot\! \lceil T_{in}^{(m)} /{2}\rceil \,. \]
\endgroup
Assume the PDP pipeline can accommodate at most $p$ CAP-Units where $p\geq 1$, then the number of recirculations required is $R = \lceil U/p \rceil$. Considering $C$ is the maximum of the all layers' input/output channels or features ($C\geq 2$), we have
\begingroup
\setlength{\abovedisplayskip}{4pt}
\setlength{\belowdisplayskip}{4pt}
\begin{equation*}
\begin{aligned}
     R = \lceil U/p \rceil &\leq \lceil (\!\sum_{n=1}^{L_{conv}} \!C^2\cdot \lceil  T/{2^n}\rceil\!+\!\sum_{m=1}^{L_{fc}}C\! \cdot \! \lceil{C} /2\rceil)/p\rceil \\
     &\leq \lceil {(T \cdot C^2 \!+ \!L_{conv} \cdot C^2 \!+\!{L_{fc} \cdot ({C^2}/{2}\! + \! C))}/{p} \rceil} \\
     &\leq \left\lceil {(T + L_{conv} + L_{fc}) \cdot C^2}/p \right\rceil \,.
\end{aligned}
\end{equation*}
\endgroup

In the worst case, the pipeline can only deploy one CAP-Unit, i.e., $p=1$. Therefore, 
$R \leq \left\lceil {(T + L_{conv} + L_{fc}) \cdot C^2} \right\rceil$.
\end{proof}

This allows us to accurately determine the number of recirculations required for the entire CNN, which is useful when deploying \quark in resource-constrained PDP hardware.
%
%

\subsubsection{Header Bits Allocation}
\label{sec:header_bit}
Since metadata is initialized to zero during recirculation, \quark stores the input and hidden layers outputs in the header. However, the header size is constrained, so \quark reuses the allocated bit positions in the header.
The outputs \({q}_{a}\) computed by a CAP-Unit layer lose their utility after serving as inputs \({q}_{x}\) for the subsequent CAP-Unit layer. Thus, they can be overlaid by other outputs. 

For convolutional layers, we need to allocate bits positions to a layer and the first group in the subsequent layer:
\begingroup
\setlength{\abovedisplayskip}{5pt}
\setlength{\belowdisplayskip}{5pt}
\[Conv\_bits= (C_{out}^{(k)} \cdot \lceil {T}/{2^{k}}\rceil + C_{in}^{(k+1)}) \cdot b \,,\]
\endgroup
where $b$ represents the quantization bit levels of \quark, and $k=\argmax_{k} C_{\text{out}}^{(k)}\lceil T/2^k \rceil$.  

For fully connected layers, we need to allocate bit positions to the layer input features and output features:
\begingroup
\setlength{\abovedisplayskip}{5pt}
\setlength{\belowdisplayskip}{5pt}
\[Fc\_bits= (T_{in}^{(l)} + T_{out}^{(l)}) \cdot b \,,\]
\endgroup
where $l=\argmax_{l}(T_{in}^{(l)} + T_{out}^{(l)})$. 

Thus, the maximum number of header bits required is:
\begingroup
\setlength{\abovedisplayskip}{5pt}
\setlength{\belowdisplayskip}{5pt}
\[Header\_bits = \max(Conv\_bits, Fc\_bits) \,.\]
\endgroup



\quark efficiently manages and stores computational data within the limited header space, ensuring optimal utilization of pipeline resources. 

\section{evaluation}
We have implemented a testbed of \quark on both P4 hardware switch (with Intel Tofino ASIC) and P4 software switch (i.e., BMv2) using P4$_{16}$ language. 

\subsection{Experiment Setup}
\textbf{Testbed}: We evaluate the performance of \quark by deploying a CNN inference model on a P4 hardware switch (Flnet S9180-32X with Intel Tofino ASIC) connecting two servers (each equipped with Intel Core i7-4771 CPU @ 3.50GHz and 40GbE network interface cards). One server is the sender to replay the traffic of datasets, 
and the other server is the receiver. Links connecting the P4 switch and servers are 40Gbps fiber optics. 

In addition to testbed with hardware switch, we also build a similar network topology in Mininet \cite{mininet} consisting of one P4 software switch (BMv2) and two hosts. 



\begin{table}[tbp]
    \caption{Features and Description.}
    \centering
    \begin{tabular}{|>{\raggedright\arraybackslash}m{2cm}|>{\raggedright\arraybackslash}m{5.5cm}|}
        \hline
        \textbf{Feature Name} & \textbf{Description} \\
        \hline
        length\_max & The maximum packet length in the flow \\
        \hline
        length\_min & The minimum packet length in the flow \\
        \hline
        length\_total & The total packet length in the flow \\
        \hline
        TCP\_flag & Cumulative number of occurrences of the FIN, SYN, ACK, PSH, RST, ECE flag \\ 
        \hline
        IAT & Inter-arrival time between two adjacent packets \\
        \hline
    \end{tabular}
    \label{tab:features}
\vspace{-15pt}
\end{table}

\textbf{Datasets and Usecases}: We utilize publicly available datasets containing network traffic traces to train and test the CNN model. The features of the first eight packets extracted from datasets are listed in Table \ref{tab:features}.

\begin{figure*}
    \subfloat[Impact of different pruning rate.]{
        \includegraphics[width=0.24\linewidth]{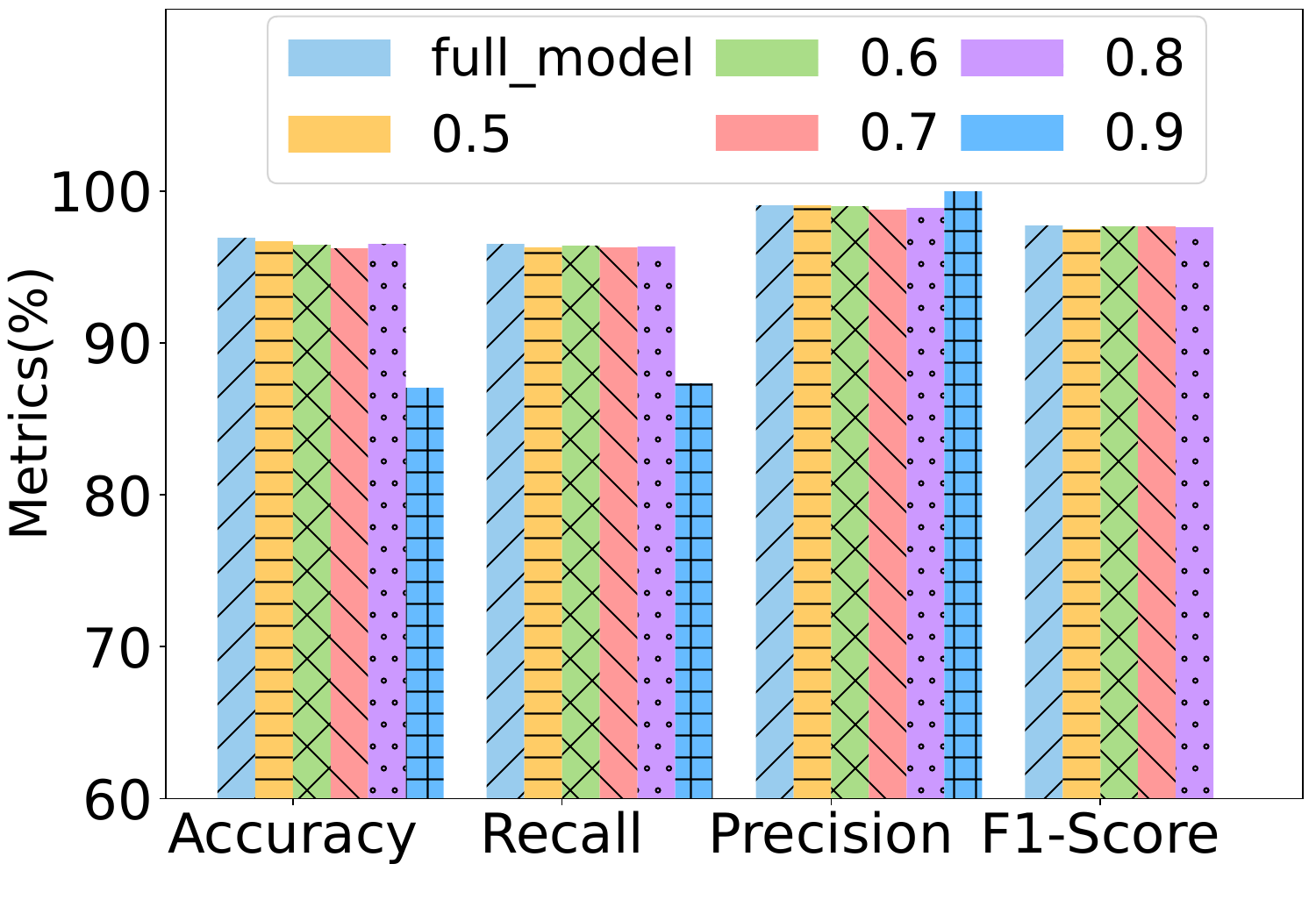}
        \label{fig:pruning}
        }
    \subfloat[FLOPs of different pruning rate.]{
        \includegraphics[width=0.24\linewidth]{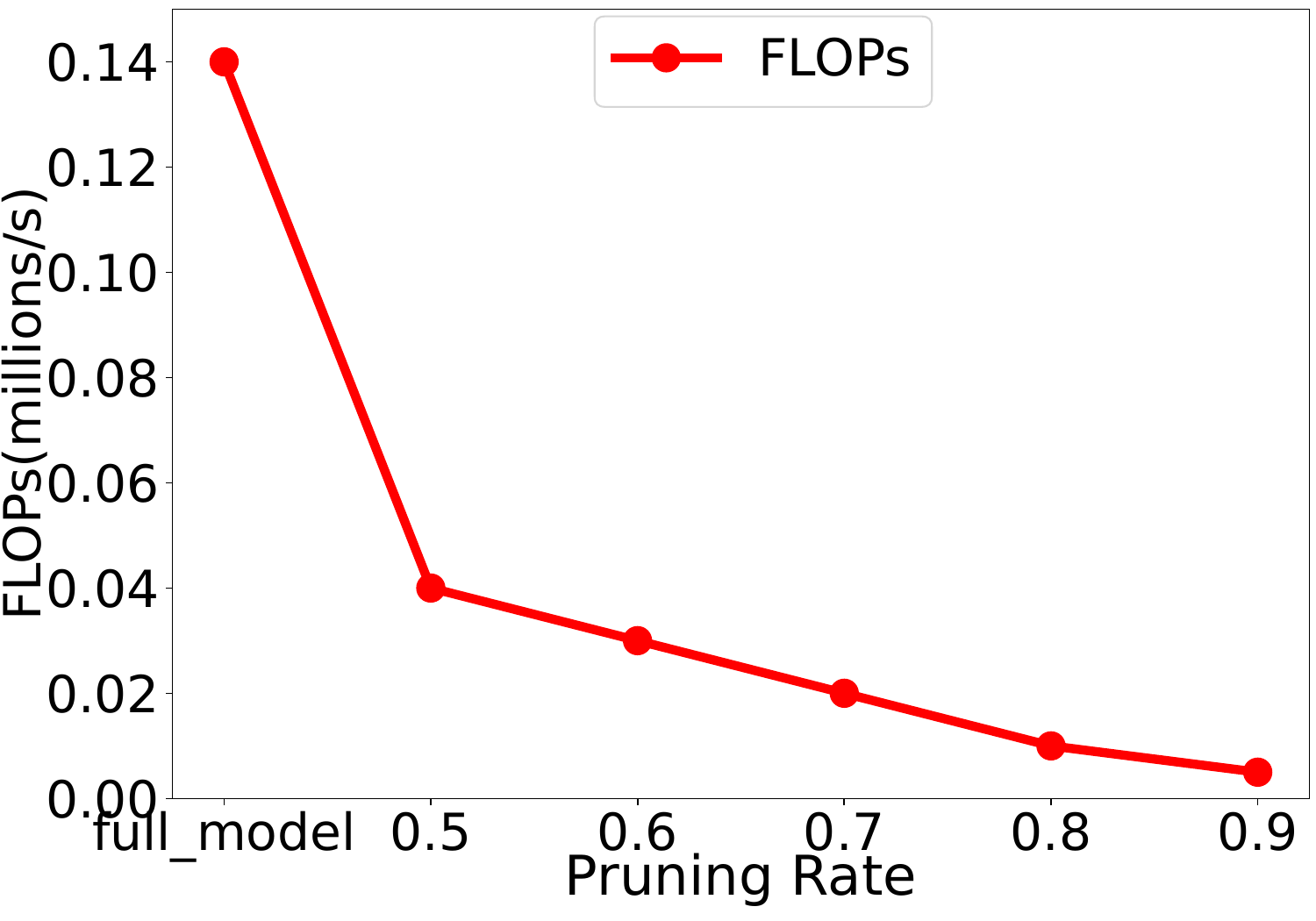}
        \label{fig:FLOPS}
    }
    \subfloat[Impact of different quantization bit.]{
        \includegraphics[width=0.24\linewidth]{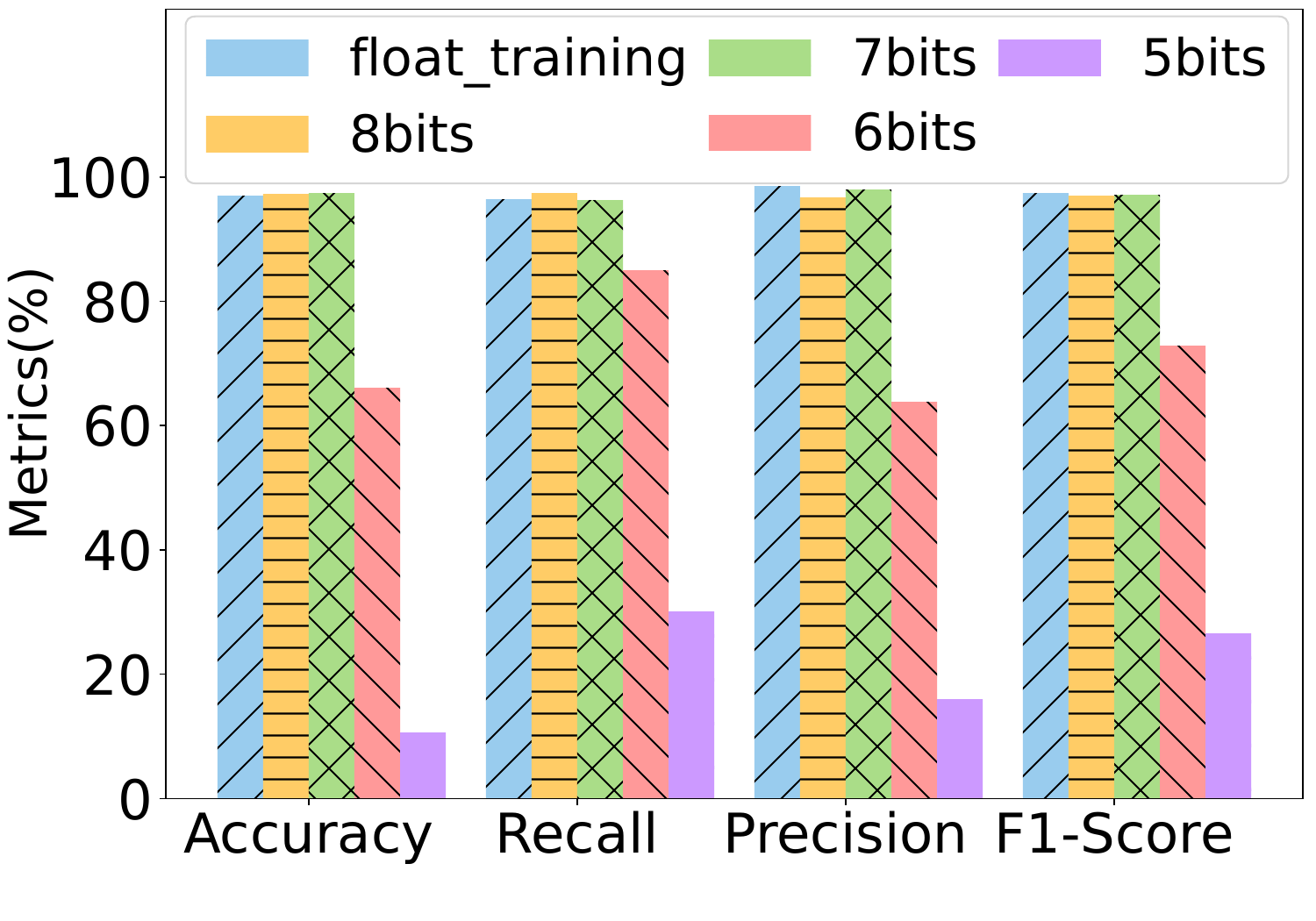}
        \label{fig:quantization}
        }
    \subfloat[Metrics of anomaly detection.]{
        \includegraphics[width=0.24\linewidth]{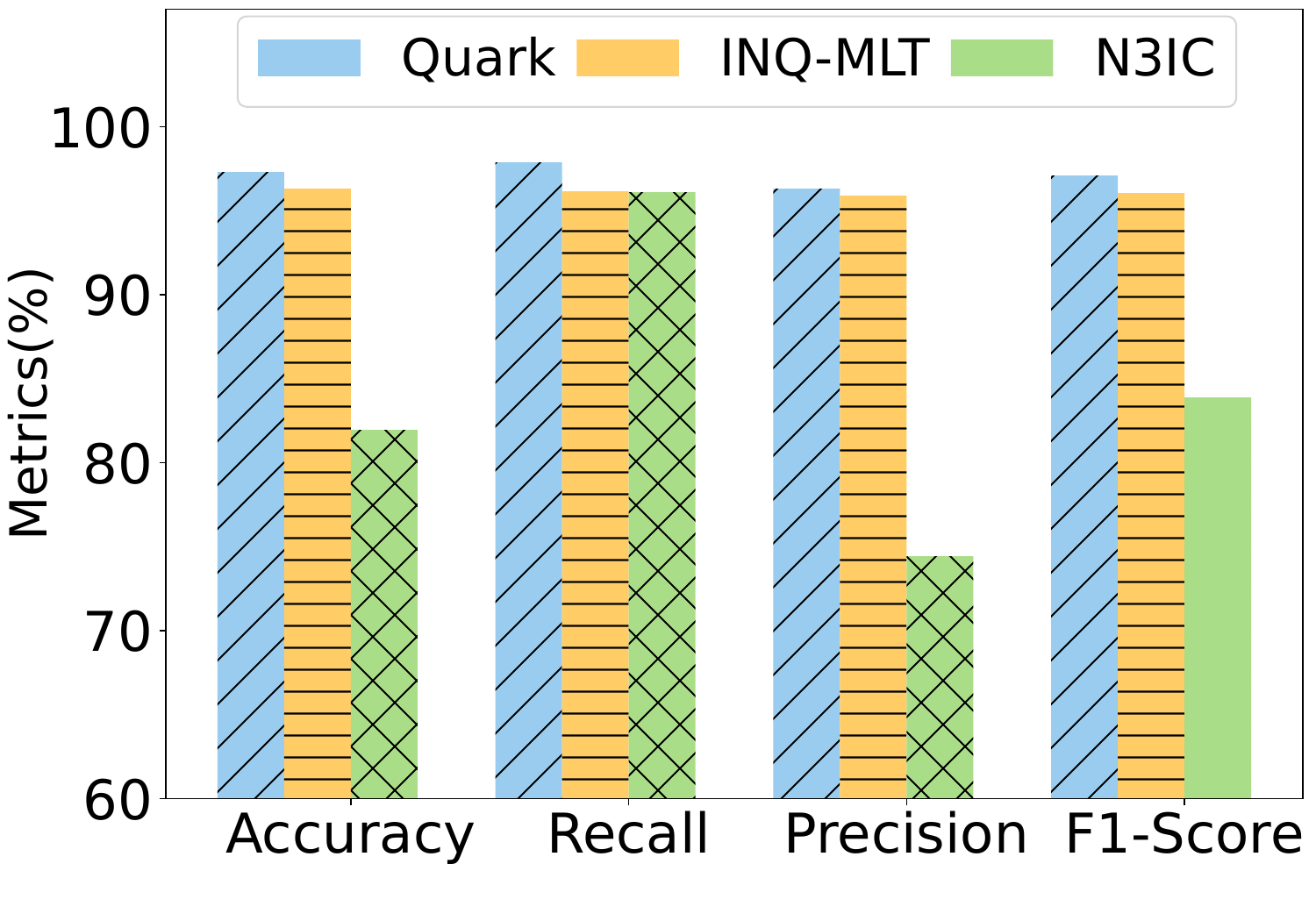}
        \label{fig:compareDection}
    }
    \vspace{5pt}
    \caption{Model performance of \quark on anomaly detection.}
    \vspace{-15pt}
\end{figure*}

\begin{itemize}
    \item \textit{Anomaly Detection}: ISCX Botnet Dataset \cite{ISCX} collects heterogeneous botnet and malware traffic alongside non-malicious traffic from real-world scenarios. We extract the features from the TCP and UDP flows of this dataset, classifying the traffic into \textit{Benign} and \textit{Malicious} categories. The training and test sets are divided according to the official guidelines provided by ISCX.
    
    \item \textit{Flow Classification}: CICIDS-2017 dataset \cite{CICIDS} includes various types of network traffic and attack scenarios. We extract flow features from the pcap files and classified the flows into \textit{Benign},\textit{ DDoS}, \textit{Patator} and \textit{PortScan} categories. Undersampling is used to handle the imbalance of the dataset. The dataset is split into 60\% training, 20\% validation, and 20\% testing subsets.
\end{itemize}

\textbf{Model Training}: We train the CNN model on a GPU server equipped with an NVIDIA GeForce RTX 2080 Ti. The CNN model consists of three convolutional layers (\({c}_{1}\)=16, \({c}_{2}\)=16, \({c}_{3}\)=16) and two fully connected layers (\({l}_{1}\)=16, \({l}_{2}\)=15). Each convolutional layer is followed by a ReLU layer and a maxpooling layer, and each fully connected layer is followed by a ReLU layer.


\textbf{Comparison Schemes}: We compare \quark with N3IC \cite{N3IC} and INQ-MLT \cite{INQ-MLT}. 
Since N3IC is designed on NetFPGA using BNN model, we only evaluate its model performance on control plane. INQ-MLT deploys CNN model on software switches and we have implemented it on BMv2 in our evaluation. Both N3IC and INQ-MLT can not be implemented on P4 hardware switches (with Intel Tofino ASIC) due to the restriction of hardware pipeline. 


\subsection{Impact of Pruning Rate and Quantization Bit Level}
We first conduct two experiments on anomaly detection to evaluate model performance of \quark. The first experiment, we fix the model size and vary the pruning rates to observe their impact on model performance. The second experiment, we fix the model size and pruning rate, then vary the quantization bit levels to assess their effect on model performance.

\textbf{Pruning Rate}:
As shown in Fig. \ref{fig:pruning}, pruning rates between 0.5 and 0.8 result in less than a 1\% decrease in all metrics compared to the original pre-compressed model, while the number of floating-point operations per second (FLOPs) decreases from 0.14M to 0.01M (shown in Fig. \ref{fig:FLOPS}). This trade-off, sacrificing less than 1\% of performance for a 92.9\% reduction in computation (at a pruning rate of 0.8), is both acceptable and beneficial for deploying large models on PDP hardware.
However, when the pruning rate reaches 0.9, the recall increases to 99.99\% and other metrics drop significantly. This indicates the model predicts almost all samples as positive due to the severely reduced number of channels after pruning. Therefore, the pruning rate should not be set too high. 

\textbf{Quantization Bit}:
We set the pruning rate to 0.8 as the baseline for floating-point training. As shown in Fig. \ref{fig:quantization}, reducing the quantization to 7 bits only causes a slight performance drop (less than 1\%) but significantly reduces memory usage, which is crucial for the limited SRAM in PDP hardware pipeline.
However, reducing the quantization to 6 bits or lower leads to severe performance degradation. At 5 bits, accuracy drops to 10.6\%, which is unacceptable.

In conclusion, selecting an appropriate pruning rate and quantization bit level can effectively conserve memory resources while preserving model performance. 
In the following experiments, \quark uses a pruning rate of 0.8 and 7 bits quantization for model compression.

\subsection{Model Performance Comparison} 
We compare the model performance of \quark with N3IC \cite{N3IC} and INQ-MLT \cite{INQ-MLT} on a server (i.e., control plane). For fairness, we use the same training data for all methods, and select the largest binary MLP model [128, 64, 10] for N3IC and the same 1D-CNN model for \quark and INQ-MLT. 

The evaluation results for the two tasks are shown in Fig. \ref{fig:compareDection} and Table \ref{tab:compareClass}. In anomaly detection, \quark performs the best among the three schemes, with an F1-Score improvement of 0.130 over N3IC and a slight increase of 0.010 over INQ-MLT. In the more complex flow classification scenario, \quark also demonstrates an average F1-Score improvement of 0.130 compared to N3IC. Both \quark and INQ-MLT exhibit distinct advantages in four classifications, with no significant difference in overall F1-Score.

The inferior performance of N3IC is primarily attributed to the accuracy loss caused by binarizing the model weights. The performance of INQ-MLT and \quark align with our assessments of the pruning rate's impact (see Fig. \ref{fig:pruning}). \quark utilizes a pruning rate of 0.8, which significantly reduces the model size while maintaining superior performance.



\begin{table*}[!t]
\centering
\renewcommand\arraystretch{1}
\caption{Performance for flow classification.}
\setlength{\tabcolsep}{4.5mm}{
\begin{tabular}{c|ccc|ccl|ccl}
\hline 
\textbf{Method}  & \multicolumn{3}{c|}{\quark}   & \multicolumn{3}{c|}{N3IC}                    & \multicolumn{3}{c}{INQ-MLT}         \\ \hline
\textbf{Metrics} & Precision & Recall & F1    & Precision & Recall & \multicolumn{1}{c|}{F1} & Precision & Recall & \multicolumn{1}{c}{F1} \\ \hline
Benign           & 0.931     & 0.942  & 0.935 & 0.783     & 0.832  & 0.807                   & 0.918     & 0.951  & 0.934                  \\
DDoS             & 0.619     & 0.625  & 0.615 & 0.494     & 0.463  & 0.478                   & 0.637     & 0.612  & 0.624                  \\
Patator          & 0.563     & 0.635  & 0.593 & 0.307     & 0.434  & 0.360                   & 0.541     & 0.642  & 0.587                  \\
PortScan         & 0.597     & 0.886  & 0.706 & 0.631     & 0.746  & 0.684                   & 0.612     & 0.901  & 0.729                  \\ \hline
Overall F1       & \multicolumn{3}{c|}{0.712} &           & 0.582  &                         &           & 0.718  &                        \\ \hline
\end{tabular}}
\label{tab:compareClass}
\vspace{-15pt}
\end{table*}
\begin{figure}
    \begin{minipage}[t]{0.49\linewidth}
        \centering
        \includegraphics[width=1\linewidth]{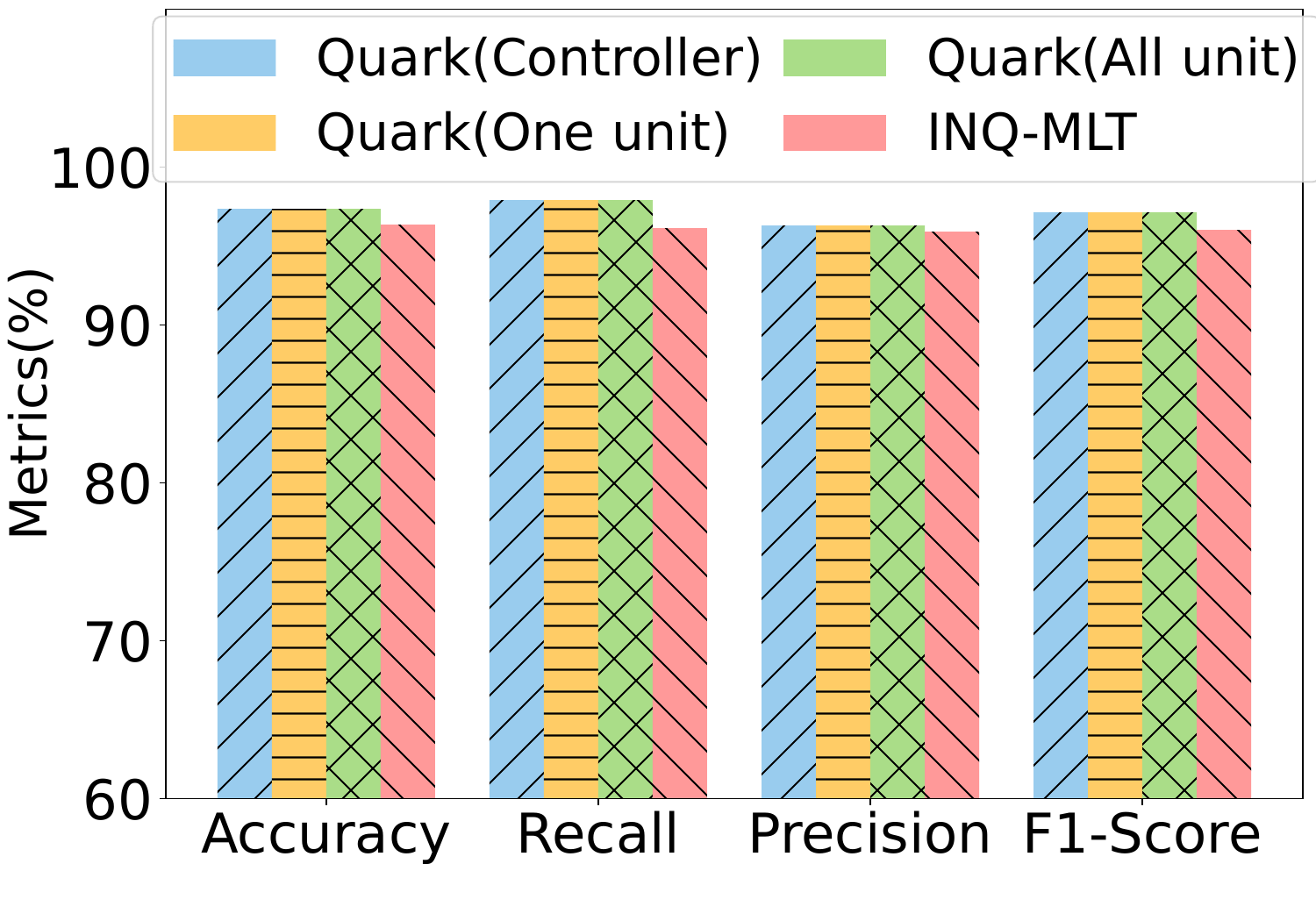}
        \vspace{-15pt}
        \caption{Anomaly detection metrics on P4 software switch.}
        \label{fig:BMv2acc}\vspace{5pt}
    \end{minipage}
    \begin{minipage}[t]{0.49\linewidth}
        \centering
        \includegraphics[width=1\linewidth]{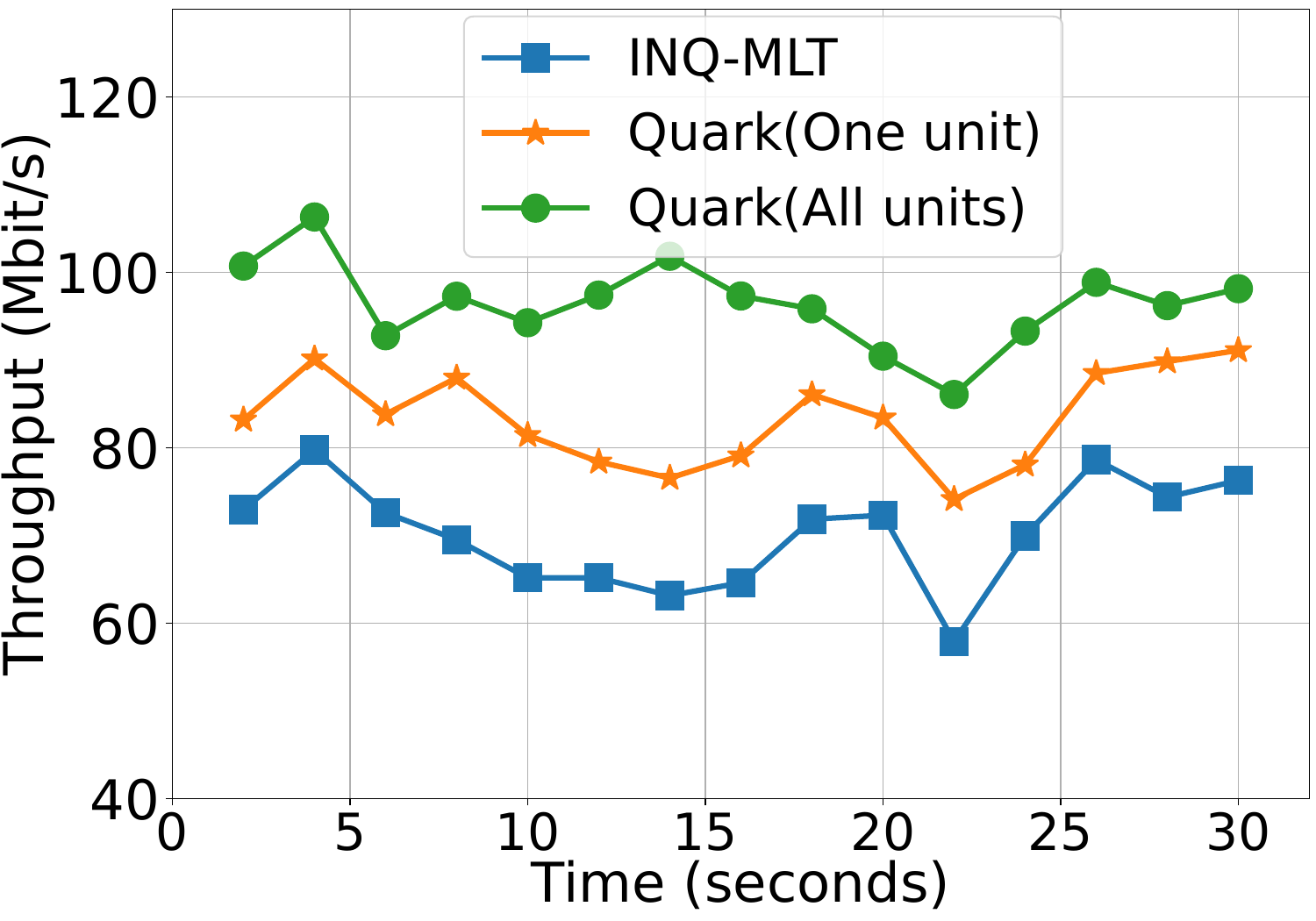}
        \vspace{-15pt}
        \caption{Throughput on P4 software switch.}
        \label{fig:BMv2thro}
    \end{minipage}
    \vspace{-15pt}
\end{figure}
\subsection{Performance in the P4 Software Switch}
We implement and compare \quark with INQ-MLT on BMv2. For \quark, only one CAP-Unit is deployed in the pipeline of BMv2. 
Additionally, we propose deploying as many units as possible within a single pipeline. Given the ample resources of BMv2, \quark offers an alternative scheme to deploy all units within a single pipeline.
We first verify whether deploying \quark to BMv2 had any impact on model performance. As shown in Fig. \ref{fig:BMv2acc}, the performance of \quark is consistent on both controller and BMv2 (PDP pipeline). 

\begin{figure}
    \vspace{-4pt}
    \begin{minipage}[t]{0.49\linewidth}
        \centering
        \includegraphics[width=1\linewidth]{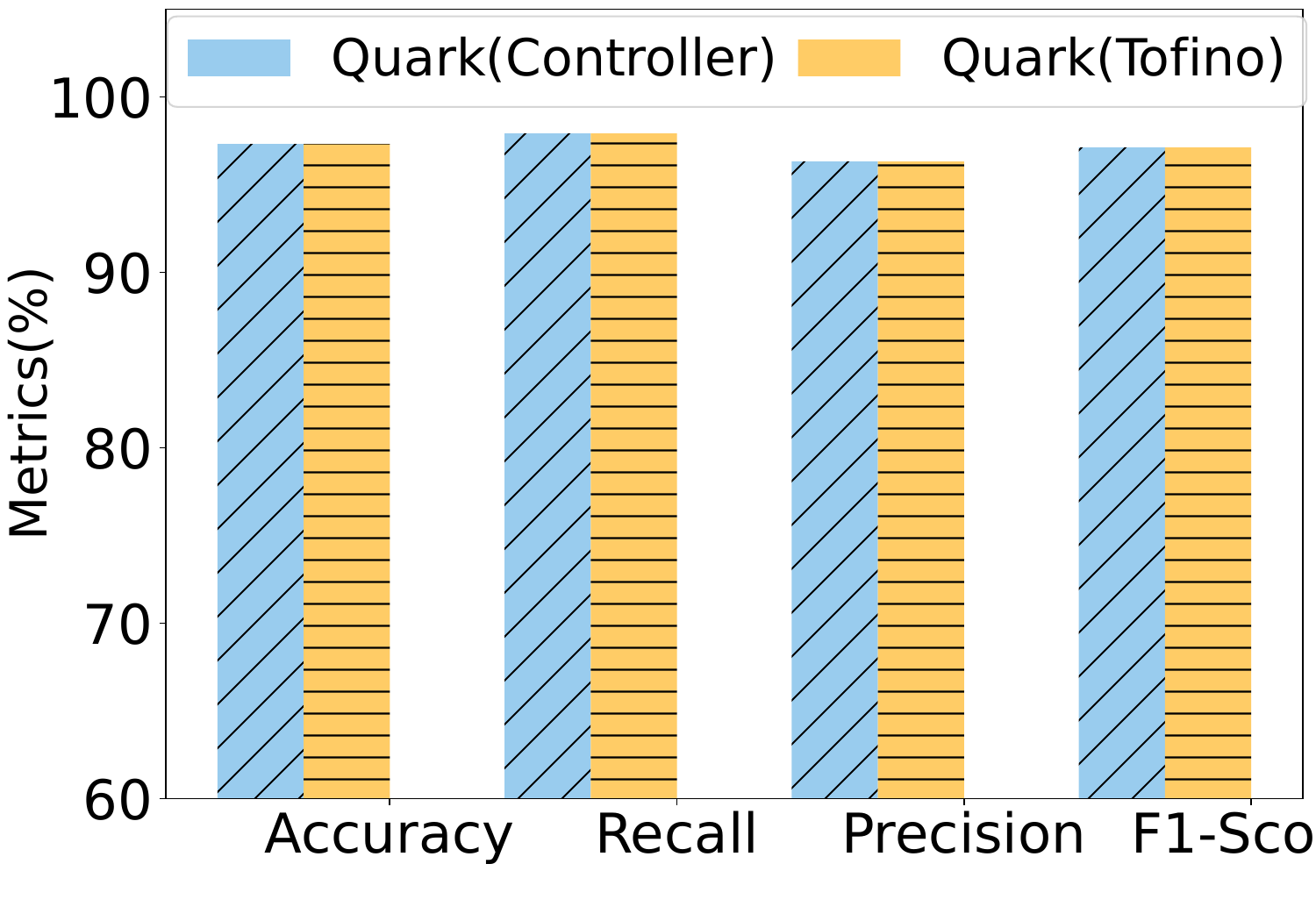}
        \vspace{-15pt}
        \caption{Anomaly detection metrics on P4 hardware switch (Tofino).}
        \label{fig:hardacc}\vspace{5pt}
    \end{minipage}
    \begin{minipage}[t]{0.49\linewidth}
        \centering
        \includegraphics[width=1\linewidth]{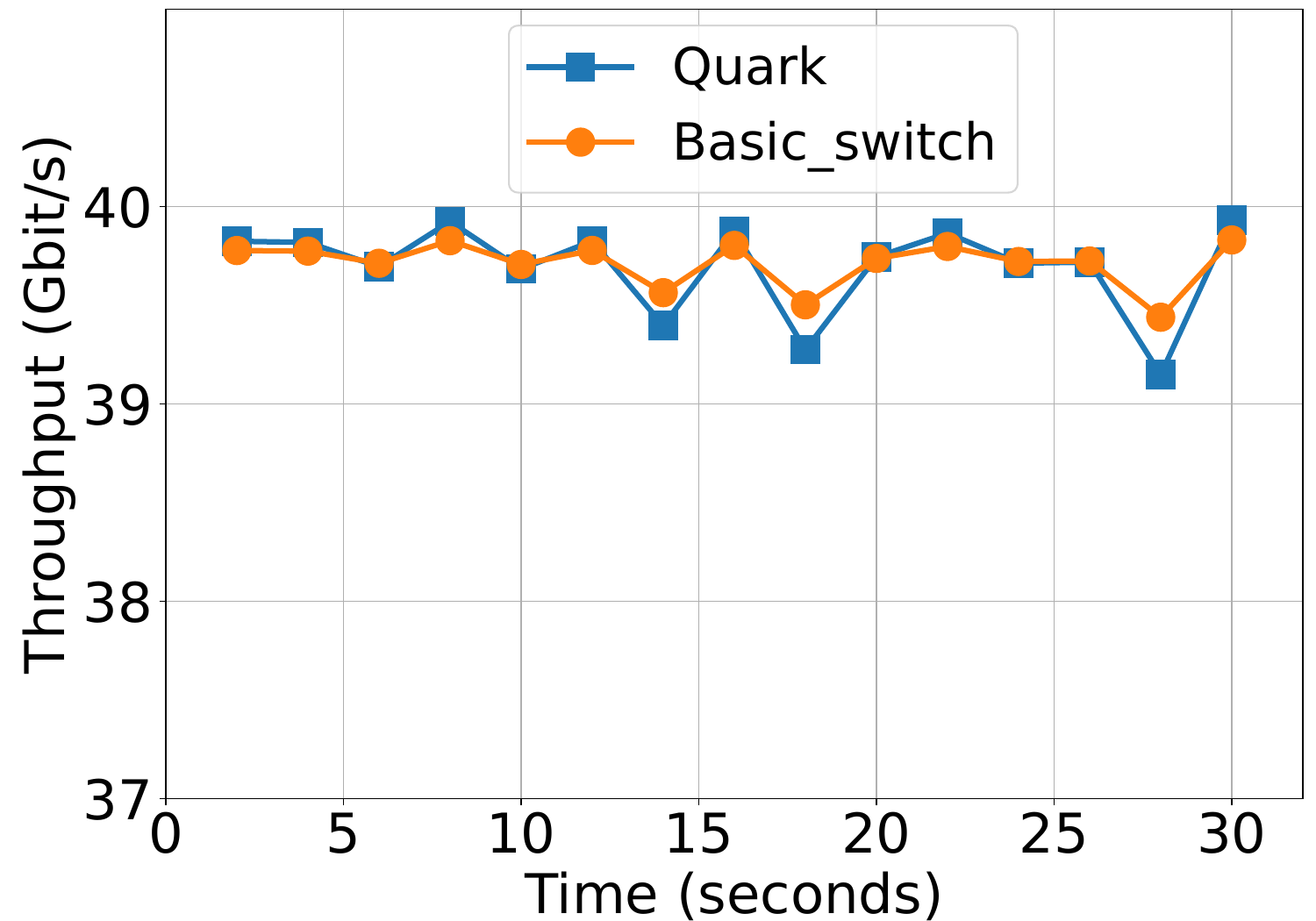}
        \vspace{-15pt}
        \caption{Throughput on P4 hardware switch (Tofino) with 40Gbps fibers.}
        \label{fig:tofinothro}        
    \end{minipage}
    \vspace{-15pt}
\end{figure}

\textbf{Comparison with INQ-MLT}: As shown in Fig. \ref{fig:BMv2acc}, \quark demonstrates approximately a 1\% improvement in model performance. However, benefiting from its pruning module, \quark only utilizes 20\% of the model size (with a pruning rate of 0.8) and 7.1\% of the computations (see Fig. \ref{fig:FLOPS}) of INQ-MLT, resulting in a significant throughput increase. As illustrated in Fig. \ref{fig:BMv2thro}, \quark achieves an average throughput improvement of 18.8\% over INQ-MLT.

\textbf{Comparison with All Units per Pipeline}: As shown in Fig. \ref{fig:BMv2acc}, both \quark solutions exhibit consistent model performance. In terms of throughput (see Fig. \ref{fig:BMv2thro}), the solution deploying all units per pipeline shows an improvement of 15.6\%. Compared to the INQ-MLT scheme, there is a significant 37.3\% increase in throughput.
This indicates that deploying as many units as possible in the pipeline effectively enhances forwarding performance.
However, due to resource constraints in Tofino, only one unit can be deployed per pipeline.





\subsection{Performance in the P4 Hardware Switch}
We have successfully implement one CAP-Unit in the P4 pipeline of Tofino hardware switch. With 102 recirculations, \quark deploys the compressed CNN model, utilizing a pruning rate of 0.8 and 7-bit quantization. 

\textbf{Accuracy}: As shown in Fig. \ref{fig:hardacc}, \quark's prediction metrics on Intel Tofino ASIC are consistent with those of the controller, achieving a 0.971 F1-Score and a 97.3\% accuracy in anomaly detection.

\textbf{Throughput}: To evaluate \quark's throughput, we implement a simple switch forwarding application, called basic\_switch, as a baseline for comparison. As shown in Fig. \ref{fig:tofinothro}, \quark achieves 39.696 Gbps throughput, only 0.04\% below the baseline's 39.712 Gbps, demonstrating line-rate operation.

\begin{figure}[t]
    \label{verifycluster}
    \centering
    \vspace{-15pt}
    \subfloat[Inference latency under different numbers of concurrent flows.]{
        \includegraphics[width=0.49\linewidth]{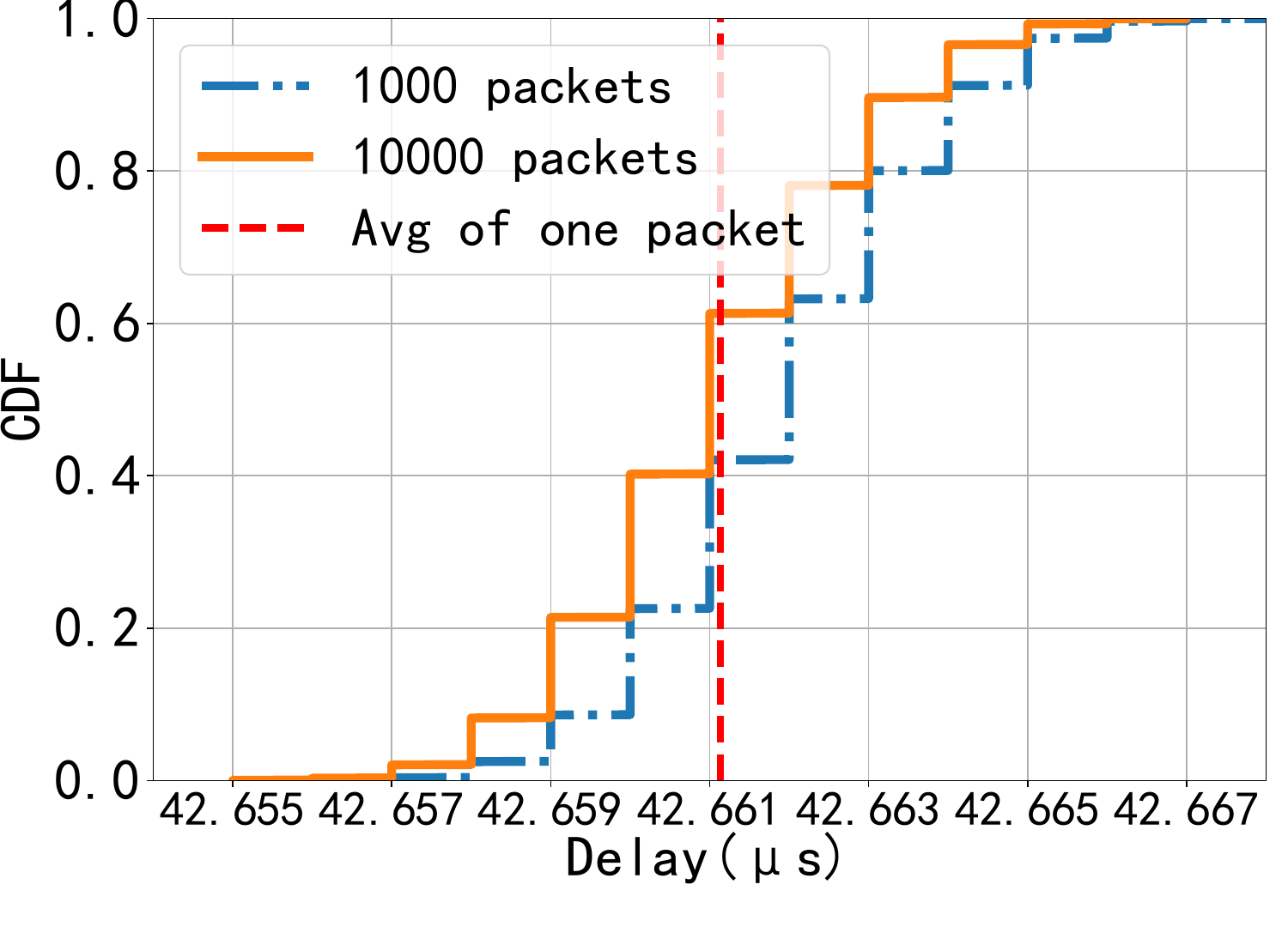}
        \label{fig:delay}
    }
    \subfloat[The number of concurrent flows.]{
        \includegraphics[width=0.49\linewidth]{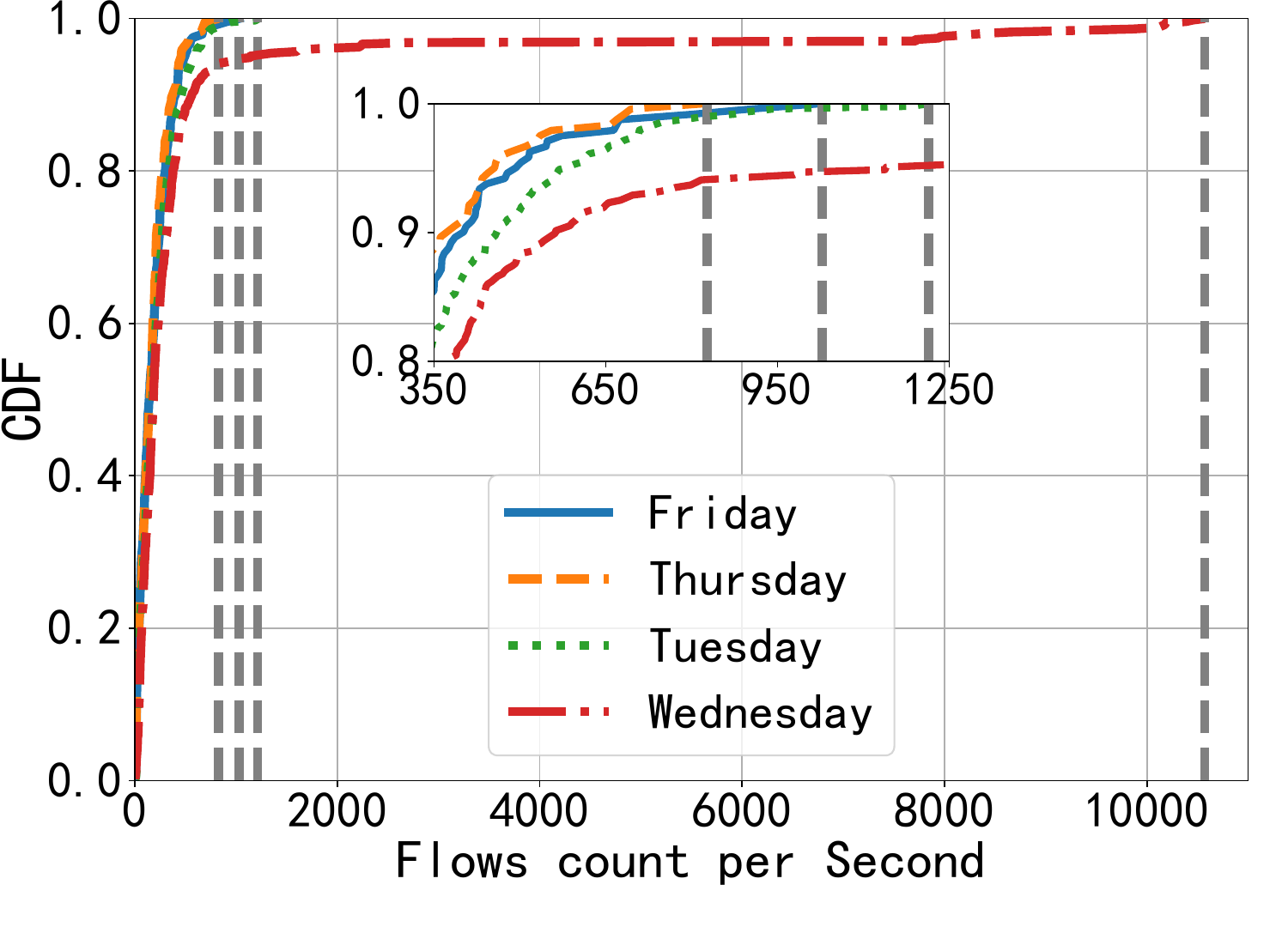}
        \label{fig:flow_num}
    }
    \vspace{2pt}
    \caption{Inference latency for CNN on P4 hardware switch.}
    \vspace{-5pt}
\end{figure}

\textbf{Inference Latency}:
The inference latency refers the delay from the start of inference when a packet enters the switch to the completion of inference and packet forwarding. 
We evaluate both single and concurrent inference latency, with the CDF results shown in Fig. \ref{fig:delay}.
For single inference, an average latency of 42.66$\mu s$ is observed from 1000 individual tests (see red dashed line). 

Analysis of the CICIDS dataset reveals that, in most cases, approximately 1000 flows arrive simultaneously within one second, with a maximum of 10571 flows on Wednesday (see Fig. \ref{fig:flow_num}).
In the concurrent flow test, the blue dotted line represents 1000 concurrent inference latency, showing an average latency of 42.66$\mu s$ with fluctuations under 0.01$\mu s$. 
The orange solid line shows similar performance for 10000 concurrent inferences. These evaluations demonstrate that \quark maintains the same performance as with a single flow, even under high concurrency conditions.

\textbf{Resource efficiency}:
Table \ref{tab:resource} shows the hardware resource consumption of \quark on Tofino switch. 
In terms of memory resources, MATs of \quark entirely rely on exact matching, which does not consume TCAM or Map RAM resources, leaving 75.73\% of the SRAM available. Additionally, 75.00\% of the table ID resources in Tofino ASIC remain unused.
Furthermore, \quark consumes only 13.6\% of the PHV bits.
Overall, the evaluation results demonstrate that \quark only consumes small portions of resources on Tofino ASIC, 
allowing the switch to concurrently deploy many other functionalities.

\begin{table}[]
\vspace{-5pt}
\caption{Hardware resource consumption on Intel Tofino ASIC.}
\label{tab:resource}
\renewcommand\arraystretch{1}

\setlength{\tabcolsep}{1mm}
\begin{tabular}{cccccc}
\hline
\textbf{Computational} & eMatch xBar & tMatch xBar & Gateway & VLIW    & Hash bits \\ \hline
Usage                  & 3.26\%      & 0.00\%      & 13.02\% & 20.83\% & 14.20\%   \\ \hline
\end{tabular}

\vspace{1mm} 

\setlength{\tabcolsep}{2.73mm}
\begin{tabular}{cccccc}
\hline
\textbf{Memory} & SRAM    & TCAM & Map RAM & TableID & Stash  \\ \hline
Usage           & 24.27\% & 0.00\% & 0.00\%  & 25.00\% & 9.38\% \\ \hline
\end{tabular}

\vspace{1mm} 

\setlength{\tabcolsep}{2.52mm}
\begin{tabular}{ccccc}
\hline
\textbf{PHV} & 8bits used  & 16bits used & 32bits used & overall bits used \\ \hline
Usage                   & 8.01\% & 27.50\% & 4.69\%  & 13.60\%   \\ \hline
\end{tabular}
\vspace{-15pt}
\end{table}

\section{Conclusion}
In this paper, we proposed \quark to enable CNN inference to be deployed entirely at line speed on the programmable hardware data plane. \quark incorporates model compression to reduce network size while maintaining high accuracy and applies quantization to address floating-point limitations on PISA switches. 
Additionally, \quark modularizes CNNs into unit-based combinations to optimize pipeline resource utilization and enable deployment on PDP
Experimental results demonstrate that \quark achieves high inference accuracy with microsecond-level latency at line rate. The authors have provided public access to their code at \href{https://github.com/AntLab-Repo/Quark-CNN-P4}{\textcolor{blue}{\underline{\textit{https://github.com/AntLab-Repo/Quark-CNN-P4}}}}.

\section*{Acknowledgement}
This work is partially supported by National Natural Science Foundation of China (NSFC) under Grant 62172189; the Innovate UK grants 10098627 and 10106629.
\clearpage

\bibliographystyle{IEEEtran} 
\bibliography{IEEEabrv,Mybib}

\begin{thebibliography}{10}
\providecommand{\url}[1]{#1}
\csname url@samestyle\endcsname
\providecommand{\newblock}{\relax}
\providecommand{\bibinfo}[2]{#2}
\providecommand{\BIBentrySTDinterwordspacing}{\spaceskip=0pt\relax}
\providecommand{\BIBentryALTinterwordstretchfactor}{4}
\providecommand{\BIBentryALTinterwordspacing}{\spaceskip=\fontdimen2\font plus
\BIBentryALTinterwordstretchfactor\fontdimen3\font minus \fontdimen4\font\relax}
\providecommand{\BIBforeignlanguage}[2]{{%
\expandafter\ifx\csname l@#1\endcsname\relax
\typeout{** WARNING: IEEEtran.bst: No hyphenation pattern has been}%
\typeout{** loaded for the language `#1'. Using the pattern for}%
\typeout{** the default language instead.}%
\else
\language=\csname l@#1\endcsname
\fi
#2}}
\providecommand{\BIBdecl}{\relax}
\BIBdecl

\bibitem{pdpsurvey}
O.~Michel, R.~Bifulco, G.~R\'{e}tv\'{a}ri, and S.~Schmid, ``The programmable data plane: Abstractions, architectures, algorithms, and applications,'' \emph{ACM Computing Surveys (CSUR)}, vol.~54, no.~4, pp. 1--36, 2021.

\bibitem{IDPs_survey}
W.-X. Liu, C.~Liang, Y.~Cui, J.~Cai, and J.-M. Luo, ``Programmable data plane intelligence: Advances, opportunities, and challenges,'' \emph{IEEE Network}, vol.~37, no.~5, pp. 122--128, 2023.

\bibitem{N3IC}
G.~Siracusano, S.~Galea, D.~Sanvito, M.~Malekzadeh, G.~Antichi, P.~Costa, H.~Haddadi, and R.~Bifulco, ``Re-architecting traffic analysis with neural network interface cards,'' in \emph{19th USENIX Symposium on Networked Systems Design and Implementation (NSDI)}, 2022, pp. 513--533.

\bibitem{swamy2022taurus}
T.~Swamy, A.~Rucker, M.~Shahbaz, I.~Gaur, and K.~Olukotun, ``Taurus: A data plane architecture for per-packet ml,'' in \emph{Proceedings of the 27th ACM International Conference on Architectural Support for Programming Languages and Operating Systems}, 2022, pp. 1099--1114.

\bibitem{BNN2}
Q.~Qin, K.~Poularakis, K.~K. Leung, and L.~Tassiulas, ``Line-speed and scalable intrusion detection at the network edge via federated learning,'' in \emph{IFIP Networking Conference (Networking)}, 2020, pp. 352--360.

\bibitem{FlowLens}
D.~Barradas, N.~Santos, L.~Rodrigues, S.~Signorello, F.~M.~V. Ramos, and A.~Madeira, ``Flowlens: Enabling efficient flow classification for ml-based network security applications,'' in \emph{Network and Distributed System Security Symposium (NDSS)}, 2021.

\bibitem{congestion}
T.~Mai, S.~Garg, H.~Yao, J.~Nie, G.~Kaddoum, and Z.~Xiong, ``In-network intelligence control: Toward a self-driving networking architecture,'' \emph{IEEE Network}, vol.~35, no.~2, pp. 53--59, 2021.

\bibitem{MLpdpSurvey}
R.~Parizotto, B.~L. Coelho, D.~C. Nunes, I.~Haque, and A.~Schaeffer-Filho, ``Offloading machine learning to programmable data planes: A systematic survey,'' \emph{ACM Computing Surveys (CSUR)}, vol.~56, no.~1, pp. 1--34, 2023.

\bibitem{DT1}
X.~Zhang, L.~Cui, F.~P. Tso, and W.~Jia, ``{pHeavy}: Predicting heavy flows in the programmable data plane,'' \emph{IEEE Transactions on Network and Service Management}, vol.~18, no.~4, pp. 4353--4364, 2021.

\bibitem{DT2}
B.~M. Xavier, R.~S. Guimarães, G.~Comarela, and M.~Martinello, ``Programmable switches for in-networking classification,'' in \emph{IEEE Conference on Computer Communications (IEEE INFOCOM)}, 2021, pp. 1--10.

\bibitem{BNN1}
G.~Siracusano and R.~Bifulco, ``In-network neural networks,'' \emph{arXiv preprint arXiv: 1801.05731}, 2018.

\bibitem{iisy2}
Z.~Xiong and N.~Zilberman, ``Do switches dream of machine learning? {Toward} in-network classification,'' in \emph{Proceedings of the 18th ACM Workshop on Hot Topics in Networks}, 2019, p. 25–33.

\bibitem{zheng2022iisy}
C.~Zheng, Z.~Xiong, T.~T. Bui, S.~Kaupmees, R.~Bensoussane, A.~Bernabeu, S.~Vargaftik, Y.~Ben-Itzhak, and N.~Zilberman, ``{IIsy}: Practical in-network classification,'' \emph{arXiv preprint arXiv:2205.08243}, 2022.

\bibitem{cnnSurvey2}
Z.~Li, F.~Liu, W.~Yang, S.~Peng, and J.~Zhou, ``A survey of convolutional neural networks: Analysis, applications, and prospects,'' \emph{IEEE Transactions on Neural Networks and Learning Systems}, vol.~33, no.~12, pp. 6999--7019, 2022.

\bibitem{INQ-MLT}
K.~Zhang, N.~Samaan, and A.~Karmouch, ``A machine learning-based toolbox for p4 programmable data-planes,'' \emph{IEEE Transactions on Network and Service Management}, 2024.

\bibitem{NetNN}
K.~Razavi, S.~D. Fard, G.~Karlos, V.~Nigade, M.~M{\"u}hlh{\"a}user, and L.~Wang, ``{NetNN}: Neural intrusion detection system in programmable networks,'' in \emph{29th IEEE Symposium on Computers and Communications (ISCC)}, 2024.

\bibitem{banana}
D.~Sanvito, G.~Siracusano, and R.~Bifulco, ``Can the network be the ai accelerator?'' in \emph{Proceedings of the 2018 Morning Workshop on In-Network Computing}, 2018, pp. 20--25.

\bibitem{Tofino_PISA}
``Intel {Tofino} switch {ASIC},'' 2024, \url{https://www.intel.com/content/www/us/en/products/details/network-io/intelligent-fabric-processors/tofino.html}, accessed on Jul. 30, 2024.

\bibitem{float_cal}
P.~Cui, H.~Pan, Z.~Li, J.~Wu, S.~Zhang, X.~Yang, H.~Guan, and G.~Xie, ``{NetFC}: Enabling accurate floating-point arithmetic on programmable switches,'' in \emph{IEEE 29th International Conference on Network Protocols (ICNP)}, 2021, pp. 1--11.

\bibitem{RNN}
J.~Yan, H.~Xu, Z.~Liu, Q.~Li, K.~Xu, M.~Xu, and J.~Wu, ``{Brain-on-Switch}: Towards advanced intelligent network data plane via {NN-Driven} traffic analysis at line-speed,'' in \emph{21st USENIX Symposium on Networked Systems Design and Implementation (NSDI)}, 2024, pp. 419--440.

\bibitem{in3}
X.~Zhang, L.~Cui, F.~P. Tso, W.~Li, and W.~Jia, ``{IN3}: A framework for in-network computation of neural networks in the programmable data plane,'' \emph{IEEE Communications Magazine}, vol.~62, no.~4, pp. 96--102, 2024.

\bibitem{defense}
M.~C. Luizelli, R.~Canofre, A.~F. Lorenzon, F.~D. Rossi, W.~Cordeiro, and O.~M. Caicedo, ``In-network neural networks: Challenges and opportunities for innovation,'' \emph{IEEE Network}, vol.~35, no.~6, pp. 68--74, 2021.

\bibitem{intrusion_detection}
R.~Vinayakumar, M.~Alazab, K.~P. Soman, P.~Poornachandran, A.~Al-Nemrat, and S.~Venkatraman, ``Deep learning approach for intelligent intrusion detection system,'' \emph{IEEE Access}, vol.~7, pp. 41\,525--41\,550, 2019.

\bibitem{traffic_classification}
W.~Wang, M.~Zhu, J.~Wang, X.~Zeng, and Z.~Yang, ``End-to-end encrypted traffic classification with one-dimensional convolution neural networks,'' in \emph{IEEE international conference on intelligence and security informatics (ISI)}, 2017, pp. 43--48.

\bibitem{zhang2023dapper}
X.~Zhang, L.~Cui, F.~P. Tso, Z.~Li, and W.~Jia, ``Dapper: Deploying service function chains in the programmable data plane via deep reinforcement learning,'' \emph{IEEE Transactions on Services Computing}, vol.~16, no.~4, pp. 2532--2544, 2023.

\bibitem{FPGA}
S.~Gandhare and B.~Karthikeyan, ``Survey on {FPGA} architecture and recent applications,'' in \emph{International Conference on Vision Towards Emerging Trends in Communication and Networking (ViTECoN)}, 2019, pp. 1--4.

\bibitem{NP}
G.~Gadre, S.~Badhe, and K.~Kulkarni, ``Network processor - {A} simplified approach for transport layer offloading on nic,'' in \emph{2016 International Conference on Advances in Computing, Communications and Informatics (ICACCI)}, 2016, pp. 2542--2548.

\bibitem{BNNsurvey}
H.~Qin, R.~Gong, X.~Liu, X.~Bai, J.~Song, and N.~Sebe, ``Binary neural networks: A survey,'' \emph{Pattern Recognition}, vol. 105, p. 107281, 2020.

\bibitem{bmv2}
``p4lang/behavioral-model,'' \url{https://github.com/p4lang/behavioral-model}, accessed on Jul. 30, 2024.

\bibitem{QAT}
M.~Nagel, M.~Fournarakis, R.~A. Amjad, Y.~Bondarenko, M.~van Baalen, and T.~Blankevoort, ``A white paper on neural network quantization,'' \emph{arXiv preprint arXiv:2106.08295}, 2021.

\bibitem{pruning}
T.~Liang, J.~Glossner, L.~Wang, S.~Shi, and X.~Zhang, ``Pruning and quantization for deep neural network acceleration: A survey,'' \emph{Neurocomputing}, vol. 461, pp. 370--403, 2021.

\bibitem{ste}
P.~Yin, J.~Lyu, S.~Zhang, S.~J. Osher, Y.~Qi, and J.~Xin, ``Understanding straight-through estimator in training activation quantized neural nets,'' in \emph{International Conference on Learning Representations (ICLR)}, 2019.

\bibitem{mininet}
``mininet/mininet,'' \url{https://github.com/mininet/mininet}, accessed on Jul. 30, 2024.

\bibitem{ISCX}
E.~Biglar~Beigi, H.~Hadian~Jazi, N.~Stakhanova, and A.~A. Ghorbani, ``Towards effective feature selection in machine learning-based botnet detection approaches,'' in \emph{IEEE Conference on Communications and Network Security}, 2014, pp. 247--255.

\bibitem{CICIDS}
I.~Sharafaldin, A.~H. Lashkari, A.~A. Ghorbani \emph{et~al.}, ``Toward generating a new intrusion detection dataset and intrusion traffic characterization,'' in \emph{International Conference on Information Systems Security and Privacy (ICISSp)}, vol.~1, 2018, pp. 108--116.

\end{thebibliography}

\end{document}